\documentclass[conference]{IEEEtran}
\usepackage{graphicx}
\usepackage{amsmath}
\usepackage{lipsum}
\usepackage{amsthm}
\usepackage[strings]{underscore}
\usepackage{amsfonts}
\usepackage{caption}
\usepackage{siunitx}
\usepackage{amssymb}
\usepackage{subcaption}
\usepackage{placeins}
\usepackage{lettrine}
\usepackage{fancyhdr} 
\usepackage{multicol, blindtext}
\usepackage{bbm}
\usepackage{calrsfs}
\usepackage{cite}
\usepackage{color}
\usepackage{cuted}
\usepackage[mathscr]{euscript}
\usepackage{multicol}
\usepackage{siunitx}
\usepackage[breaklinks=true]{hyperref}
\usepackage{algorithm}
\usepackage{algpseudocode}
\usepackage{authblk}
\makeatletter
\newcommand{\Rmnum}[1]{\expandafter\@slowromancap\romannumeral #1@}

\theoremstyle{plain}
\newtheorem{theorem}{Theorem}

\newtheorem{proposition}{Proposition}

\newtheorem{lemma}{Lemma}
\theoremstyle{definition}
\newtheorem{definition}{Definition}
\theoremstyle{remark}

\makeatletter
\def\BState{\State\hskip-\ALG@thistlm}
\makeatother
\ifCLASSINFOpdf
\else
\fi

\begin{document}
\title{Status Updates with Priorities: Lexicographic Optimality}
\author[*]{Ali Maatouk}
\author[$\S$]{Yin Sun}
\author[$\dagger$]{Anthony Ephremides}
\author[*]{Mohamad Assaad}
\affil[*]{TCL Chair on 5G, Laboratoire des Signaux et Syst\`emes, CentraleSup\'elec, Gif-sur-Yvette, France }
\affil[$\S$]{Department of ECE, Auburn University, Auburn,
AL 36849}
\affil[$\dagger$]{ECE Dept., University of Maryland, College Park, MD 20742}
\maketitle
\thispagestyle{fancy}
\pagestyle{fancy}
\fancyhf{}
\fancyheadoffset{0cm}
\renewcommand{\headrulewidth}{0pt} 
 \captionsetup[subfigure]{width=0.9\textwidth}
\renewcommand{\footrulewidth}{0pt}
\renewcommand*{\thepage}{\scriptsize{\arabic{page}}}
\fancyhead[R]{\thepage}
\fancypagestyle{plain}{%
   \fancyhf{}%
   \fancyhead[R]{\thepage}%
}
\begin{abstract}
%In this paper, we consider $N$ information streams sharing a common service facility. Based on how crucial their data are, we assume that the information streams are divided into $I$ priority classes. With that in mind, we introduce the notion of lex-age-optimality that captures both the age-optimality and the order of time-cruciality
%between the streams in a general multi-class scheduling scenario. 
%Equipped with this definition, we propose an online scheduling policy which we show to be lex-age-optimal for minimizing any time-dependent,
%symmetric, and non-decreasing penalty function of the
%ages of the streams in a stochastic ordering sense. Lastly, we present numerical results that
%corroborate the theoretical findings and highlight the importance of this notion of lex-age-optimality. 
In this paper, we consider a transmission scheduling problem, in which several streams of status update packets with diverse priority levels are sent through a shared channel to their destinations. We introduce a notion of \emph{Lexicographic age optimality}, or simply \emph{lex-age-optimality},  to evaluate the performance of multi-class status update policies. In particular, a lex-age-optimal scheduling policy first minimizes the Age of Information (AoI) metrics for high-priority streams, and then, within the set of optimal policies for high-priority streams, achieves the minimum AoI metrics for low-priority streams. We propose a new scheduling policy named Preemptive Priority, Maximum Age First, Last-Generated, First-Served (PP-MAF-LGFS), and prove that the PP-MAF-LGFS scheduling policy is lex-age-optimal. This result holds (i) for minimizing any time-dependent, symmetric, and non-decreasing age penalty function; (ii) for minimizing any non-decreasing functional of the stochastic process formed by the age penalty function; and (iii) for the cases where different priority classes have distinct arrival traffic patterns, age penalty functions, and age penalty functionals. For example, the PP-MAF-LGFS scheduling policy is lex-age-optimal for minimizing the mean peak age of a high-priority stream and the time-average age of a low-priority stream. Numerical results are provided to illustrate our theoretical findings.
%
%
%
%
%
%
%
%
%
%In this paper, we consider multiple information streams divided into several priority classes and sharing a common service facility. We introduce the notion of lex-age-optimality that captures both the age-optimality and the order of time-cruciality between the streams in a general multi-class scheduling scenario. Equipped with this definition, we propose the Preemptive Priority,
%Maximum Age First, Last-Generated, First-Served (PP-MAF-LGFS) scheduling policy, which we show to be lex-age-optimal. These lex-age-optimality results hold for (i) minimizing any time-dependent, symmetric, and non-decreasing penalty function of the ages in a stochastic ordering sense, (ii) minimizing any non-decreasing functional of the age penalty process, and (iii) the cases where each priority class has distinct traffic patterns and data timeliness requirements. Lastly, we present numerical results that corroborate the theoretical findings and highlight the importance of this notion of lex-age-optimality.
\let\thefootnote\relax\footnotetext{This work has been supported by ONR N000141812046, NSF CCF1813078, NSF CNS1551040, NSF CCF1420651, and ONR N00014-17-1-2417.}
%\relax\footnote{}
%\footnote{}
%\blfootnote{This work has been supported by ONR N000141812046, NSF CCF1813078, NSF CNS1551040, and NSF CCF1420651.}
\end{abstract}
\IEEEpeerreviewmaketitle
\section{Introduction}
\lettrine{D}{ue} to the proliferation of cheap hardware, remote monitoring has become the norm for modern technology applications. In these applications, a monitor is interested in timely updates about the status of a remote system. These status updates range from vehicles' position and velocity in autonomous driving to the temperature and humidity levels of a certain area in environmental monitoring. To capture this notion of timeliness, the Age of Information (\textbf{AoI}), which is defined as the information time lag at the monitor, has been introduced \cite{6195689}. Due to its widespread application range and its ability to quantify the freshness of information, the AoI is regarded as a fundamental performance metric in communication networks. The AoI has attracted a significant surge of interest in recent years \cite{6195689,8262777,8445981,8732378,2019arXiv190102873Z,2019arXiv191003564B,8613408,8764468,6875100,8000687,8437927,8822722,8812616,2019arXiv190706604M,8406891}. In particular, transmission scheduling of multiple update streams in both centralized and distributed settings has been explored in \cite{8406945,Kadota:2018:SPM:3311528.3311547,8845254,2020arXiv200103096M,8938128,8807257,2018arXiv180103975J,2019arXiv190100481M}. For example, the authors in \cite{8406945} proposed both age-optimal and near age-optimal scheduling policies for the single and multi-server cases, respectively. 
In a variety of real-life applications, information streams are assigned different priorities based on how crucial and time-sensitive their data are. A simple example is a vehicular network where data can be divided into two categories: crucial safety data and non-safety-related information. As the former is more time-sensitive than the latter, it should always be given a higher priority by the service facility \cite{5948952}. Accordingly, priority-based scheduling problems have been extensively studied in the queuing theory literature for different performance measures (e.g., delay, throughput). In \cite{Hajek2000LexOptimalMS}, a notion of Lexicographic optimality, or simply, lex-optimality, was introduced for throughput maximization in multi-class scheduling scenarios. The idea of lex-throughput-optimality is to first find a class of optimal scheduling policies $\Pi_{\text{opt}}$ that maximize the throughput of a high priority class, and then find the optimal scheduling policies within $\Pi_{\text{opt}}$ that maximize the throughput of the low priority class.  Therefore, it is clear how it elegantly provides high priority streams the best possible service by the facility, and at the same time, optimize the performance of the low priority streams.

%
%Among the works on multi-class scheduling environments, the notion of Lexicographic optimality has been proposed for throughput maximization in \cite{Hajek2000LexOptimalMS}. 
%This notion was proposed to guarantee the throughput optimality of the high priority classes, and then accordingly, maximize the throughput of the low priority classes. 

There exist several recent studies on status updates with multiple priority classes. In \cite{8437591}, the authors considered multiple information streams, each with a different priority, and sharing a common service facility with null or one waiting room in the queue that is shared by the streams. The authors studied the case where a high priority packet would preempt a lower priority packet, which is then discarded. Using a tool named Stochastic Hybrid Systems (\textbf{SHS}), the authors found an expression of the average age of each stream. The arrival rate of each stream was then optimized accordingly. In another work \cite{8849695}, the authors investigated the same settings of \cite{8437591} but by letting each stream have its own buffer space. Most recently, closed forms of the average Peak Age of Information (\textbf{PAoI}) were found in M/M/1/1 settings where streams are assigned different priorities\cite{2019arXiv190612278X}. As can be seen from the past works in the literature, the research efforts lay mainly in finding closed-form expressions of the average AoI/PAoI in a particular scenario, and for a specific arrival traffic model, to provide insights on the performance of the system. Accordingly, the question of what is the age-optimal scheduling policy in a multi-class priority-based scheduling scenario remains open. In our paper, we find an answer to this question. To that end, we summarize in the following the key contributions of this paper:
\begin{itemize}
\item We introduce the notion of Lexicographic optimality for the age minimization framework, which we will refer to as the lex-age-optimality. The lex-age-optimality elegantly captures both the age-optimality and the order of time-cruciality between the streams in a general multi-class scheduling scenario. This approach guarantees that the performance of low priority streams is optimized while ensuring that high priority streams are granted the best possible service by the facility. 
\item In the case of a single server with i.i.d. exponential service times, we propose the \emph{Preemptive Priority, Maximum Age First, Last-Generated, First-Served} (\textbf{PP-MAF-LGFS}) scheduling policy. Using a sample-path argument, we show that this policy is lex-age-optimal. Our lex-age-optimality results are not constrained to the traditional minimization of the average AoI and PAoI frameworks previously adopted in \cite{8437591,8849695,2019arXiv190612278X}. In fact, they hold for (i) minimizing any time-dependent, symmetric, and non-decreasing penalty function of the ages, and (ii) minimizing any non-decreasing functional of the age penalty process. We note that, unlike the previous works on multi-class status updates, our lex-age-optimality results are not bound to any traffic arrival distribution. Moreover, they hold when the priority classes have distinct traffic patterns and different dissatisfaction levels of the aged information. This showcases the wide scope of our results as classes typically represent diverse applications, each with its data timeliness requirements. For example, we could be interested in minimizing the average PAoI for a class and the average AoI for another. 
\end{itemize}
The rest of the paper is organized as follows: Section II is
dedicated to the system model where the required definitions and the queuing model are presented. In Section III, we introduce the notion of 
lex-age-optimality and propose a lex-age-optimal policy in the single exponential server settings. Numerical results that corroborate these findings are laid out in Section IV while the paper
is concluded in Section V.

%Consequently, these approaches fall short in the following four areas:
%\begin{enumerate}
%\item These approaches solely focused on studying the AoI of the system for a particular policy. The optimality of the employed policy among all the possible scheduling policies was not established. Accordingly, the question of what is the age-optimal scheduling policy in a multi-class priority-based scheduling scenario remains open.
%\item The sum AoI (or PAoI) were the sole functionals of the age process  considered in these works. Therefore, if the level of dissatisfaction with the aged information is different than these two functionals, the results become invalid.
%\item The above works have solely considered Poisson
%arrivals for all priority classes. Knowing that
%the priority classes typically have distinct, and not necessarily Poisson, traffic patterns, a more general approach is therefore required. 
%\item These efforts focused on minimizing the \emph{total} average AoI/PAoI of the network in the framework of priority-based scheduling.
%This approach fails to capture the differences in time-cruciality between the priority classes as it lets each stream contribute equally to the overall penalty of the system. 
%\end{enumerate}
%Our paper aims to address the above issues. To that end, the following are the key contributions of this paper:
\color{black}
\begin{figure*}[ht]
\centering
\includegraphics[width=.65\linewidth]{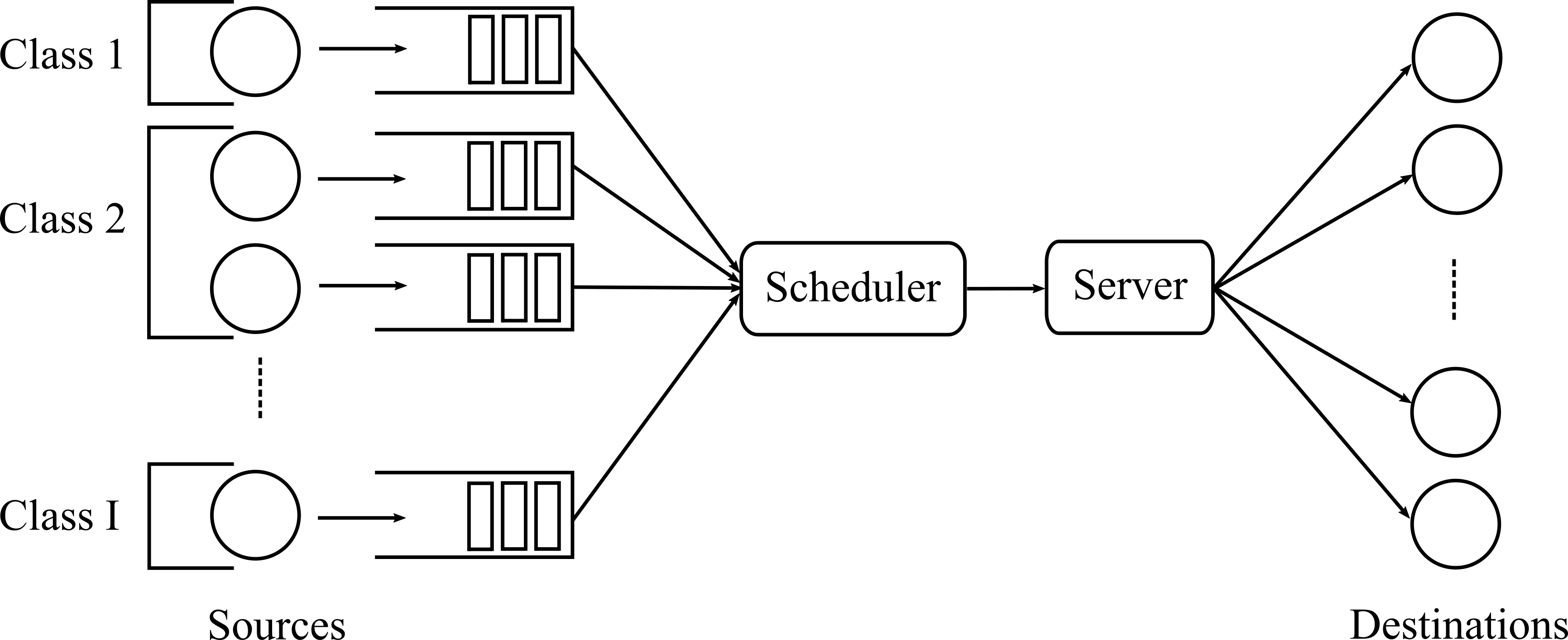}
\caption{System model.}
\vspace{-15pt}
\label{systmodel}
\end{figure*}
\section{System Model}
\label{systemmodelsec}
\subsection{Notations and Definitions}
We let $x$ and $\boldsymbol{x}$ denote deterministic scalars and vectors respectively. Similarly, we will use $X$ and $\boldsymbol{X}$ to denote random scalars and vectors respectively. Let $x^i$ denote the $i$-th element of vector $\boldsymbol{x}$, and let $x^{[i]}$ denote the $i$-th largest element of vector $\boldsymbol{x}$. Hence, $x^{[1]}$ and $x^{[N]}$ denote the largest and smallest elements of vector $\boldsymbol{x}$ respectively. We denote by $[\boldsymbol{x}]$ the sorted version of vector $\boldsymbol{x}$ (i.e. $[\boldsymbol{x}]^{i}=x^{[i]}$). Vector $\boldsymbol{x}\in\mathbb{R}^N$ is said to be smaller than $\boldsymbol{y}\in\mathbb{R}^N$, denoted by $\boldsymbol{x}\leq\boldsymbol{y}$, if $x^i\leq y^i$ for $i=1,\ldots,N$.  The composition of two functions $f$ and $g$ is denoted by $f\circ g(\boldsymbol{x})=f(g(\boldsymbol{x}))$. A function $p: \mathbb{R}^N\mapsto \mathbb{R}$ is said to be symmetric if 
$p(\boldsymbol{x})=p([\boldsymbol{x}])$ for all $\boldsymbol{x}\in \mathbb{R}^N$.
%
%
%
%
%
%
%Throughout this paper, we will refer to deterministic scalars and vectors in lowercase $x$ and $\boldsymbol{x}$ respectively. Similarly, we will refer to random scalars and vectors in uppercase $X$ and $\boldsymbol{X}$ respectively. Each vector's element will be referred to by a subscript such as $x_i$. In the sequel, we will use $x_{[i]}$ to denote the $i$-th largest element of the vector $\boldsymbol{x}$. More specifically, $x_{[1]}$ and $x_{[N]}$ will denote the largest and smallest element of any vector $\boldsymbol{x}\in\mathbb{R}^N$ respectively. Moreover, we denote by $[\boldsymbol{x}]$ the sorted version of the vector $\boldsymbol{x}$ (i.e. $[\boldsymbol{x}]_{1}=x_{[1]}$). \color{black}To compare vectors, we will use the element-wise comparison operator. To that end, we say that $\boldsymbol{x}\in\mathbb{R}^N$  is smaller than $\boldsymbol{y}\in\mathbb{R}^N$  and we denote it by $\boldsymbol{x}\leq\boldsymbol{y}$, if $x_i\leq y_i\:\: \forall i\in\{1,\ldots,N\}$. The composition of two functions $f$ and $g$, which is equal to $f(g(\boldsymbol{x}))$, will be denoted by $f\circ g(\boldsymbol{x})$.\\
%With the above notation cleared out, we provide in the following several useful definitions to our paper's theoretical analysis.
%\begin{definition}
%A function $p: \mathbb{R}^N\mapsto \mathbb{R}$ is said to be symmetric if 
%\begin{equation}
%p(x^{1},\ldots,x^{N})=p(x^{[1]},\ldots,x^{[N]})\quad \forall \boldsymbol{x}\in \mathbb{R}^N.
%\end{equation}
%\end{definition}
\noindent Next, we define stochastic ordering, which we will use in our subsequent age-optimality analysis.
%
%
%As our goal is to find age-optimal policies, we present in the following the notion of stochastic ordering which will be necessary to our optimality proofs. 
\begin{definition}\textit{Stochastic Ordering of Random Variables} \cite{2007}: 
A random variable X is said to be stochastically smaller than
a random variable Y, denoted by $X\leq_{st}Y$, if $\Pr(X>t)\leq \Pr(Y>t)\:\:\forall t\in\mathbb{R}$.
\end{definition}
%\noindent The above definition can be extended to the case of random vectors $\boldsymbol{X},\boldsymbol{Y}\in\mathbb{R}^{N}$ as follows.
\begin{definition}\textit{Stochastic Ordering of Random Vectors} \cite{2007}: A set $\mathcal{U}\subseteq\mathbb{R}^{N}$ is called upper if $\boldsymbol{y}\in\mathcal{U}$ whenever $\boldsymbol{x}\leq\boldsymbol{y}$ and $\boldsymbol{x}\in\mathcal{U}$. Let $\boldsymbol{X}$ and $\boldsymbol{Y}$ be two $n$-dimensional random vectors, $\boldsymbol{X}$ is said to be stochastically smaller than $\boldsymbol{Y}$, denoted by $\boldsymbol{X}\leq_{st}\boldsymbol{Y}$, if
\begin{equation}
\Pr(\boldsymbol{X}\in\mathcal{U})\leq \Pr(\boldsymbol{Y}\in\mathcal{U})\quad\forall \mathcal{U}\subseteq\mathbb{R}^{N}.
\end{equation}
%where $\mathcal{I}\subseteq\mathbb{R}^{N}$ is an upper set defined as follows:
%\begin{equation}
%\forall \boldsymbol{x}\in\mathcal{I}\:\: \forall\boldsymbol{y}\in \mathbb{R}^{N}: \boldsymbol{x}\leq\boldsymbol{y}\Rightarrow \boldsymbol{y}\in\mathcal{I}
%\end{equation}
\end{definition}
%\noindent Similarly, we present in the following the notion of stochastic ordering of stochastic processes.
\begin{definition}\textit{Stochastic Ordering of Stochastic Processes} \cite{2007}:
A stochastic process $\{X(t),t\geq0\}$ is said to be stochastically smaller than a stochastic process $\{Y(t),t\geq0\}$, denoted by $\{X(t),t\geq0\}\leq_{st}\{Y(t),t\geq0\}$, if for any sequence of time instants $t_1<t_2<\ldots<t_m\in\mathbb{R}^+$
\begin{equation}
(X(t_1),X(t_2),\ldots,X(t_m))\leq_{st} (Y(t_1),Y(t_2),\ldots,Y(t_m)).
\end{equation}
\end{definition}
\noindent Let $\mathbb{V}$ be the set of Lebesgue measurable functions on $[0,\infty)$, i.e.,
\begin{equation}
\mathbb{V}=\{g: [0,\infty)\mapsto \mathbb{R} \textnormal{ is Lebesgue measurable}\}.
\label{defineV}
\end{equation}
A functional $\phi: \mathbb{V}\mapsto \mathbb{R}$ is said to be non-decreasing if $\phi(g_1)\leq\phi(g_2)$ holds for all $g_1,g_2\in\mathbb{V}$ that satisfy $g_1(t)\leq g_2(t)$ for $t\in[0,\infty)$. We note that $\{X(t),t\geq0\}\leq_{st}\{Y(t),t\geq0\}$ if, and only if, \cite{2007}
\begin{equation}
\mathbb{E}[\phi(\{X(t),t\geq0\})]\leq\mathbb{E}[\phi(\{Y(t),t\geq0\})]
\label{stochasticordrr}
\end{equation}
holds for every non-decreasing functional $\phi$ for which the expectations in (\ref{stochasticordrr}) exist.
\subsection{Queuing Model}
Consider the status-update system illustrated in Fig. \ref{systmodel}, where $N$ streams of update packets are sent through a common service facility. Each update stream has a buffer space, which can be infinite or finite. The server can process at most one packet at a time. The packet service times are i.i.d. across streams and time. The information streams are divided into $I$ priority classes, with streams of the same class $i$ having the same priority. Each information stream is indexed by two components $(i,j)$, where $i$ denotes the class index and $j$ denotes the stream index within class $i$. The classes are indexed
in a decreasing order of priority. In other words, classes
$1$ and $I$ are the highest and lowest priority classes,
respectively. Let $J_i$ be the number of steams in class $i$. Let $s_{i,j}$ and $d_{i,j}$ denote the source and destination nodes of stream $(i,j)$, respectively. Different streams can have different source and/or destination nodes.
%
%
%
%
%
%
%
%
%
%
%
%With the above division in mind, we will be characterizing each information stream by two components $(i,j)$ where:
%\begin{itemize}
%\item $i$ is the class index to which the stream belongs. Without loss of generality, we suppose that the classes are indexed in a decreasing order of priority. In other words, classes $1$ and $I$ are the highest and lowest priority classes respectively.
%\item $j$ is the index of the stream within the class $i$. We let $J_i$ be the total number of streams of class $i$.
%\end{itemize}

The system starts operating at time $t=0$. The $n$-th update packet of stream $(i,j)$ is generated at time $S^{i,j}_n$, arrives to the stream's buffer at time $A^{i,j}_n$, and is delivered to the destination $d_{i,j}$ at time $D^{i,j}_n$. Accordingly, we always have $0\leq S^{i,j}_1\leq S^{i,j}_2\leq\ldots$ and $S^{i,j}_n\leq A^{i,j}_n\leq D^{i,j}_n$. We consider in our paper 
the following class of synchronized packet generation and arrival processes.
\begin{definition}\textit{Intra-class Synchronized Sampling and Arrivals}:
The packet generation and arrival times are said to be synchronized across streams within each class, if for all classes $i=1,\ldots,I$, there exist two sequences $\{S_1^i,S_2^i,\ldots\}$ and $\{A_1^i,A_2^i,\ldots\}$ such that for all $n=1,2,\ldots$ and $j=1,\ldots,J_i$
\begin{equation}
S^{i,j}_n=S_n^i, \quad A^{i,j}_n=A_n^i.
\end{equation}
\end{definition}
%
%
%
% 
%
%Throughout this paper, we focus on the family of \emph{intra-class} synchronized packets generation and arrival processes. More specifically, we suppose that:
%\begin{equation}
%\Upsilon^{i,j}=\Upsilon^{i}\quad \forall j\in\{1,\ldots,J_i\}
%\end{equation}
Note that we let each class have its unique traffic pattern as we do not impose \emph{inter-class} synchronization. In practice, the synchronization between streams within each class can take place when these streams are synchronized by the same clock, e.g., in monitoring and control applications \cite{277716,1316761}. An example of such scenario is a vehicular network where safety-related data (e.g., position and velocity) are generated every $T$ time units, while other data of lower priority can have different traffic pattern (e.g., updates on the traffic are generated every $T'$ time units) \cite{5948952}. Also, when $J_i=1$ for a certain class $i$, the synchronization assumption within class $i$ reduces to arbitrary packet generation and arrival processes for the aforementioned class. It is worth mentioning that our work is not restricted to any traffic arrival distribution, and can include arbitrary arrival processes where packets may arrive out of order of their generation times. \color{black}In the sequel, we let 
\begin{equation}
\mathcal{I}=\{(S_n^i,A_n^i),\: i=1,\ldots,I, \quad  n=1,2,\ldots \}
\end{equation}
denote the sequence of generation/arrival times for all the classes of the system. We suppose that $\mathcal{I}$ is independent of the service times of the packets and is not altered by the choice of the scheduling policy. 

Let $\pi$ represents a scheduling policy that determines the packets being sent over time. Let $\Pi$ denotes the
set of all \emph{causal} scheduling policies, i.e., where the decisions are taken without any knowledge of the future. A policy is said to be work-conserving if the service facility is kept busy whenever there exist one or more unserved packet in the queues. We let $\Pi_{wc}$ denote the set of work-conserving causal policies. A policy is said to be preemptive if it allows the service facility to switch to transmitting another packet at any time.
%
%
%Within the set $\Pi$, we define the following subsets of scheduling policies:
%\begin{itemize}
%\item \emph{Work-conserving policies $\Pi_{wc}$}: 
%\item \emph{Preemptive policies $\Pi_{p}$}: 
%\item \emph{Non-preemptive policies}: A policy is said to be non-preemptive if the switch to transmitting another packet by the server at any time is not allowed.
%\end{itemize}
\subsection{Age Penalty Functions and Functionals}
We define the instantaneous age of information of stream $(i,j)$ at time instant $t$ as:
\begin{equation}
\Delta^{i,j}\small(t\small)=t-\max\{S^{i,j}_n:D^{i,j}_n\leq t,\:\:n=1,2,\ldots\},
\end{equation}
which is the difference between the current time $t$ and the generation time of the freshest
packet that has been delivered to the destination $d_{i,j}$. We let $\boldsymbol{\Delta}^{i}(t)=(\Delta^{i,1}(t),\ldots,\Delta^{i,{J_i}}(t))$ denote the age vector at time $t$ of all streams belonging to class $i$. Additionally, we let $\boldsymbol{\Delta}(t)=(\boldsymbol{\Delta}^{1}(t),\ldots,\boldsymbol{\Delta}^{I}(t))$ denote the age vector of all streams at time $t$.
% where $\boldsymbol{\Delta}^{i}(t)=(\Delta^{i,1}(t),\ldots,\Delta^{i,{J_i}}(t))$ is the age vector of all streams belonging to class $i$.
%
%
% Throughout this paper, we will write the age vector at each time instant $t$ as follows:
%\begin{equation}
%\boldsymbol{\Delta}(t)=(\boldsymbol{\Delta}^{1}(t),\ldots,\boldsymbol{\Delta}^{c_L}(t))
%\end{equation}

We introduce an age penalty function $p_t\circ\boldsymbol{\Delta}^{i}(t)$ that represents the level of dissatisfaction with the aged information at time $t$ for class $i$, where $p_t: \mathbb{R}^{J_i}\mapsto \mathbb{R}$ is a non-decreasing function of $\boldsymbol{\Delta}^{i}(t)$. Some commonly used age penalty functions are listed below.
%In this paper, we are interested in minimizing the average of a penalty function of the age vector over all the possible causal scheduling policies. To that end, we let $p_t\circ\boldsymbol{\Delta}^{i}(t)$ be the level of dissatisfaction with the aged information at time $t$ for class $i$, where $p_t: \mathbb{R}^N\mapsto \mathbb{R}$ is a non-decreasing function of $\boldsymbol{\Delta}^{i}(t)$. 
\begin{itemize}
\item The sum age of the $J_i$ streams:
\begin{equation}
p_{\text{sum}}\circ\boldsymbol{\Delta}^{i}(t)=\sum_{j=1}^{J_i} \Delta^{i,j}(t).
\label{functional1}
\end{equation}
\item The maximum age of the $J_i$ streams:
\begin{equation}
p_{\text{max}}\circ\boldsymbol{\Delta}^{i}(t)=\max_{j=1,\ldots,J_i}\Delta^{i,j}(t).
\label{functional2}
\end{equation}
\item The average age threshold violation of the $J_i$ streams:
\begin{equation}
p_{\text{exceed}-\alpha}\circ\boldsymbol{\Delta}^{i}(t)=\frac{1}{J_i}\sum_{j=1}^{J_i}\mathbbm{1}_{\{\Delta^{i,j}(t)>\alpha\}}.
\label{functional3}
\end{equation}
where $\mathbbm{1}_{\{.\}}$ is the indicator function, and $\alpha$ is a fixed age threshold that should not be violated.
\item The sum age penalty function of the $J_i$ streams:
\begin{equation}
p_{\text{pen}}\circ\boldsymbol{\Delta}^{i}(t)=\sum_{j=1}^{J_i} g(\Delta^{i,j}(t)),
\label{functional3}
\end{equation}
where $g:\mathbb{R}^{+}\mapsto \mathbb{R}$ is a non-decreasing function. For instance, an exponential function $g(\Delta^{i,j})=\exp(a\Delta^{i,j})$ with $a>0$ can be used for control applications where the system is vulnerable to outdated information and the need for fresh information grows quickly with respect to the age \cite{8000687}.
\end{itemize}
%
%
% This penalty function can be used In applications where
%
%This showcases the .
%
%
% [IT paper]
%
%
%For example, a stair-shape function
%g1() = bac with a  0 can be used to characterize
%the dissatisfaction of data staleness when the information
%of interests is checked periodically, and an exponential
%function g2() = ea is appropriate for online learning
%and control applications where the desire for information
%refreshing grows quickly with respect to the age [4].
%\begin{center}
%\begin{tabular}{|c|c|}
%
% \hline
% Symbol & Expression  \\
%  \hline
%  $p_{\text{sum}}$  & $\displaystyle\sum_{j=1}^{J_i} \Delta^{i,j} $ \\
%  \hline
% $p_{\text{max}}$  & $\displaystyle\max_{j}\Delta^{i,j}\:\: j=1,\ldots,J_i$ \\
% \hline
% $p_{\text{pen}}$  & $\displaystyle\sum_{j=1}^{J_i} g(\Delta^{i,j})$ \\
% & $g:\mathbb{R}^{+}\mapsto \mathbb{R}$ is a non-decreasing function\\
% \hline
%\end{tabular}
% \captionof{table}{Frequently used penalty functions}
% \label{possiblepenaltyfunctions}
%\end{center} 
We focus in our paper on the family of symmetric and non-decreasing penalty functions:
\begin{equation*}
\mathcal{P}_{\text{sym}}=\{p:[0,\infty)^{N}\mapsto\mathbb{R} \text{ is symmetric and non-decreasing}\}.
\end{equation*}
%
%This is a fairly large class of age penalty functions, where the
%function p can be discontinuous, non-convex, or non-separable.
%It is easy to see
%fpavg; pmax; pms; pl-norm; psum-penaltyg  Psym:
%
This class of penalty functions $\mathcal{P}_{\text{sym}}$ is fairly large, and include the provided age penalty functions (\ref{functional1})-(\ref{functional3}). Furthermore, we point out that $p_t$ can change over time, which represents the time-variant importance of the information streams. This highlights the generality of our considered penalty functions.

In addition to age penalty functions, we use non-decreasing functionals $\phi(\{p_t\circ\boldsymbol{\Delta}^{i}(t), t\geq0\})$ of the age penalty process $\{p_t\circ\boldsymbol{\Delta}^{i}(t), t\geq0\}$ to represent the level of dissatisfaction with the aged information of class $i$, which we dub as the age penalty functionals. Some examples of these functionals are listed below.
\begin{itemize}
\item The time-average age penalty:
\begin{equation}
\phi_{\text{avg}}(\{p_t\circ\boldsymbol{\Delta}^{i}(t), t\geq0\})=\frac{1}{T}\int_{0}^{T}p_t\circ\boldsymbol{\Delta}^{i}(t)dt.
\end{equation}
\item The average peak age penalty:
\begin{equation}
\phi_{\text{peak}}(\{p_t\circ\boldsymbol{\Delta}^{i}(t), t\geq0\})=\frac{1}{K}\sum_{k=1}^{K}A_k,
\label{peakagefunctional}
\end{equation}
where $A_k$ denotes the $k$-th peak value of $p_t\circ\boldsymbol{\Delta}^{i}(t)$ since time $t=0$. In particular, when class $i$ has only one stream and $p_t\circ \Delta^{i}(t) = \Delta^{i}(t)$, (\ref{peakagefunctional}) reduces to the widely used average peak age metric \cite{6875100,2019arXiv190612278X}.
\end{itemize}
We consider in our paper that different priority classes can have distinct age penalty functions and functionals. This is of paramount importance as each priority class typically represents a different application, each with its data timeliness requirements. For example, in a vehicular network, time-crucial safety data related to vehicle position should be delivered promptly. Typically, the system performance is affected by the peak of the maximum age of the delivered updates. Accordingly, we can choose the maximum age penalty function $p_{\text{max}}$ and the peak age penalty $\phi_{\text{peak}}$ for this class of traffic. On the other hand, updates on gas tanks levels require an average timely delivery and, consequently, we can choose the penalty function $p_{\text{sum}}$ and the time-average age penalty functional $\phi_{\text{avg}}$ for this class. This further highlights the generality of our considered framework.\color{black}

%Note that the age vector $\boldsymbol{\Delta}(t)$ depends on both time t and policy $\pi$, and the age penalty function
%$p_t$ may change over time. 
In the sequel, we use $\{\boldsymbol{\Delta}_{\pi}^{i}(t), t\geq0\}$ and $\{p_t\circ\boldsymbol{\Delta}_{\pi}^{i}(t), t\geq0\}$ to represent the stochastic age process and penalty process of class $i$ respectively when policy $\pi$ is adopted. We assume that the initial age $\boldsymbol{\Delta}_{\pi}(0^-)$ at time $t=0^-$ is the same for all $\pi\in\Pi$.\color{black}
\section{Multi-class Multi-stream Scheduling}
\subsection{Lexicographic Optimality for Age Minimization}
In the sequel, we will introduce the notion of Lexicographic optimality for the age minimization framework, which we will refer to as the lex-age-optimality. As it has been previously detailed in the introduction section, the recent efforts on age analysis in multi-class environments focused mainly on finding closed-form expressions of the average AoI/PAoI of the multi-class system for a particular policy \cite{8437591,8849695,2019arXiv190612278X}. Note that, in the multi-class system case, the minimization of the total average AoI/PAoI falls short in capturing the differences in time-cruciality between the classes. In fact, this approach lets each stream contribute equally to the penalty of the system regardless of its class. As will be seen in the following, the lex-age-optimality elegantly solves this issue and provides a new direction of age analysis in multi-class scheduling environments.
\begin{definition}\textit{Lex-age-optimality}:
A scheduling policy $P\in\Pi$ is said to be level $1$ lex-age-optimal within $\Pi$ if for all $\mathcal{I}$, $p_t\in\mathcal{P}_{\text{sym}}$ and $\pi\in \Pi$
\begin{align}
[\{p_t\circ\boldsymbol{\Delta}^{1}_{P}(t), t\geq0\}|\mathcal{I}]
\leq_{st}[\{p_t\circ\boldsymbol{\Delta}^{1}_{\pi}(t), t\geq0\}|\mathcal{I}].
\label{level1lexoptimalitydef}
\end{align}
We let $\Pi_{\text{lex-opt}}^{1}\subseteq\Pi$ denote the set of scheduling policies that are level $1$ lex-age-optimal. In addition, $P$ is said to be level $k$ lex-age-optimal for $k=2,\ldots,I$ if it is level $k-1$ lex-age-optimal, and for all $\mathcal{I}$, $p_t\in\mathcal{P}_{\text{sym}}$ and $\pi\in \Pi_{\text{lex-opt}}^{k-1}$
%
%if,
%
%
%Generally, , we say that 
%
%
%a policy $P\in\Pi$  for all $\mathcal{I}$ and $p_t\in\mathcal{P}_{\text{sym}}$, it is level $k-1$ lex-age-optimal and 
\begin{align}
[\{p_t\circ\boldsymbol{\Delta}^{k}_{P}(t), t\geq0\}|\mathcal{I}] \leq_{st} [\{p_t\circ\boldsymbol{\Delta}^{k}_{\pi}(t), t\geq0\}|\mathcal{I}],
\label{levelklexoptimalitydef}
\end{align}
where $\Pi_{\text{lex-opt}}^{k-1}$ is the set of scheduling policies that are level $k-1$ lex-age-optimal. If policy P is level $k$ lex-age-optimal simultaneously for all $k=1,\ldots, I$, it is said to be lex-age-optimal. 
\label{definitionlexoptimality}
\end{definition}
According to (\ref{stochasticordrr}), (\ref{level1lexoptimalitydef}) can be equivalently expressed as
\begin{align}
&\mathbb{E}[\phi(\{p_t\circ\boldsymbol{\Delta}^{1}_{P}(t), t\geq0\})|\mathcal{I}]\nonumber\\=&\min_{\pi\in \Pi}\mathbb{E}[\phi(\{p_t\circ\boldsymbol{\Delta}^{1}_{\pi}(t), t\geq0\})|\mathcal{I}],
\label{newdefinitionlevel1}
\end{align}
for all $\mathcal{I}$, $p_t\in\mathcal{P}_{\text{sym}}$, and non-decreasing functional $\phi:\mathbb{V}\mapsto \mathbb{R}$,
provided that the expectations in (\ref{newdefinitionlevel1}) exist. Similarly, an equivalent formulation of the level $k$ lex-age-optimality (\ref{levelklexoptimalitydef}) of a policy $P\in\Pi_{\text{lex-opt}}^{k-1}$ is 
\begin{align}
&\mathbb{E}[\phi(\{p_t\circ\boldsymbol{\Delta}^{k}_{P}(t), t\geq0\})|\mathcal{I}]\nonumber\\=&\min_{\pi\in \Pi_{\text{lex-opt}}^{k-1}}\mathbb{E}[\phi(\{p_t\circ\boldsymbol{\Delta}^{k}_{\pi}(t), t\geq0\})|\mathcal{I}],
\label{newdefinitionlevelk}
\end{align}
for all $\mathcal{I}$, $p_t\in\mathcal{P}_{\text{sym}}$, and non-decreasing functional $\phi:\mathbb{V}\mapsto \mathbb{R}$,
provided that the expectations in (\ref{newdefinitionlevelk}) exist.

The goal of the lex-age-optimality is to guarantee the age-optimality of high priority classes, and optimize the age performance of the low priority classes accordingly. To see how this is achieved, we recall from (\ref{newdefinitionlevel1}) that a level $1$ lex-age-optimal policy $P$ achieves the smallest possible expected value of any non-decreasing functional $\phi$ of the stochastic age penalty process $
[\{p_t\circ\boldsymbol{\Delta}^{1}(t), t\geq0\})|\mathcal{I}]$ among all causal policies. Next, to maintain the age-optimality of the highest priority class, our attention is restricted to the set of scheduling policies that are level $1$ lex-age-optimal. We have denoted this set by $\Pi_{\text{lex-opt}}^{1}$. To that end, and as seen in (\ref{newdefinitionlevelk}), a policy $P$ is level $2$ lex-age-optimal if it achieves the smallest possible expected value of any non-decreasing functional $\phi$ of the stochastic age penalty process $
[\{p_t\circ\boldsymbol{\Delta}^{2}(t), t\geq0\})|\mathcal{I}]$ among all level $1$ lex-age-optimal policies. This showcases how the lex-age-optimality captures the time-cruciality of streams since, by definition, lex-age-optimal policies grant high priority streams the best possible performance without being influenced by low priority streams. Then, while ensuring the age-optimality of the high priority streams, the performance of the low priority streams is optimized. 
\color{black}
\subsection{Lex-Age-Optimal Policy for Exponential Service Time
}
\begin{figure*}[ht]
\centering
\begin{subfigure}{0.3333\textwidth}
  \centering
  \includegraphics[width=.99\linewidth]{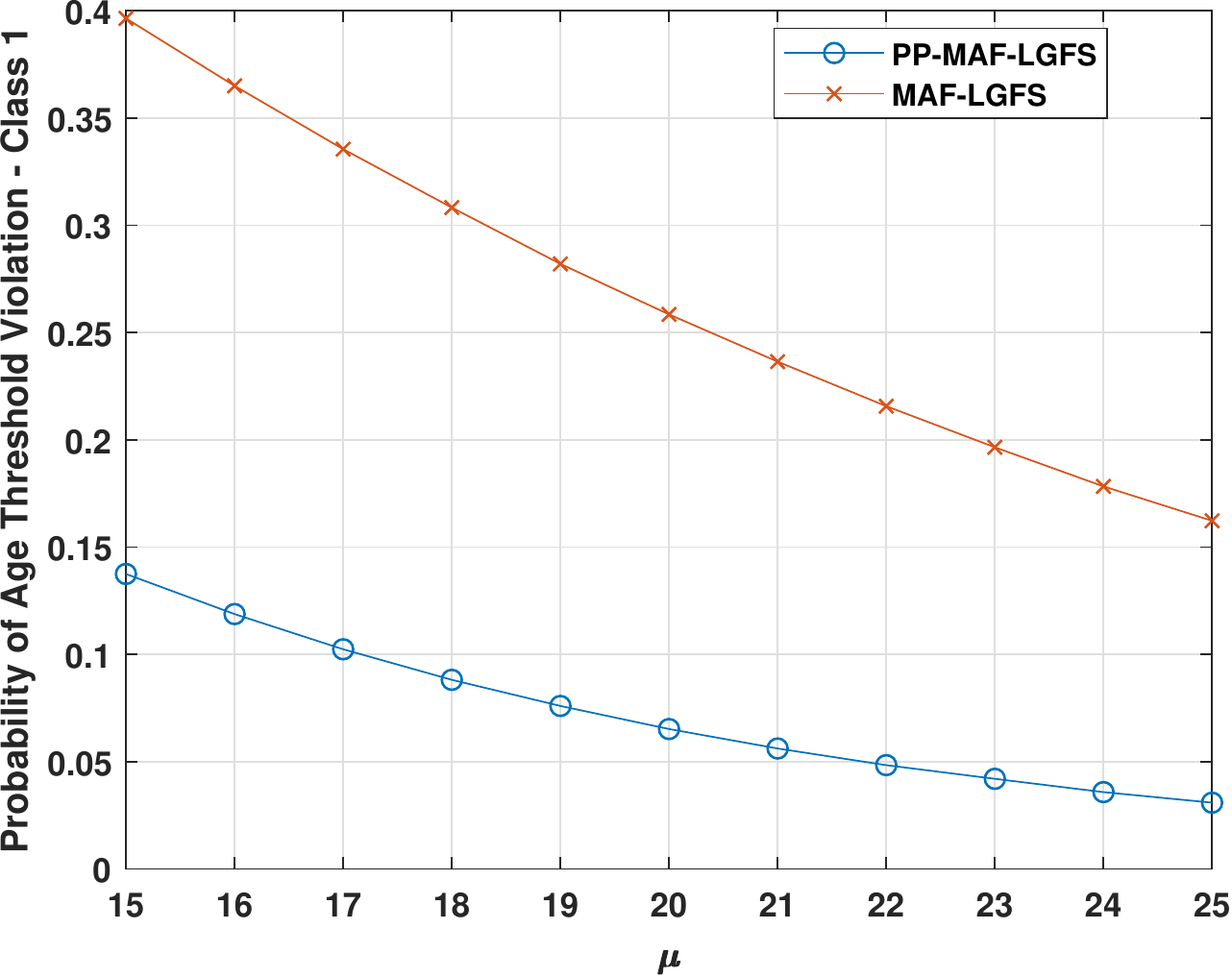}
  \caption{Probability of age violation for class $1$.}
    \label{sim1}
\end{subfigure}%    
\begin{subfigure}{0.3333\textwidth}
\centering
  \includegraphics[width=.99\linewidth]{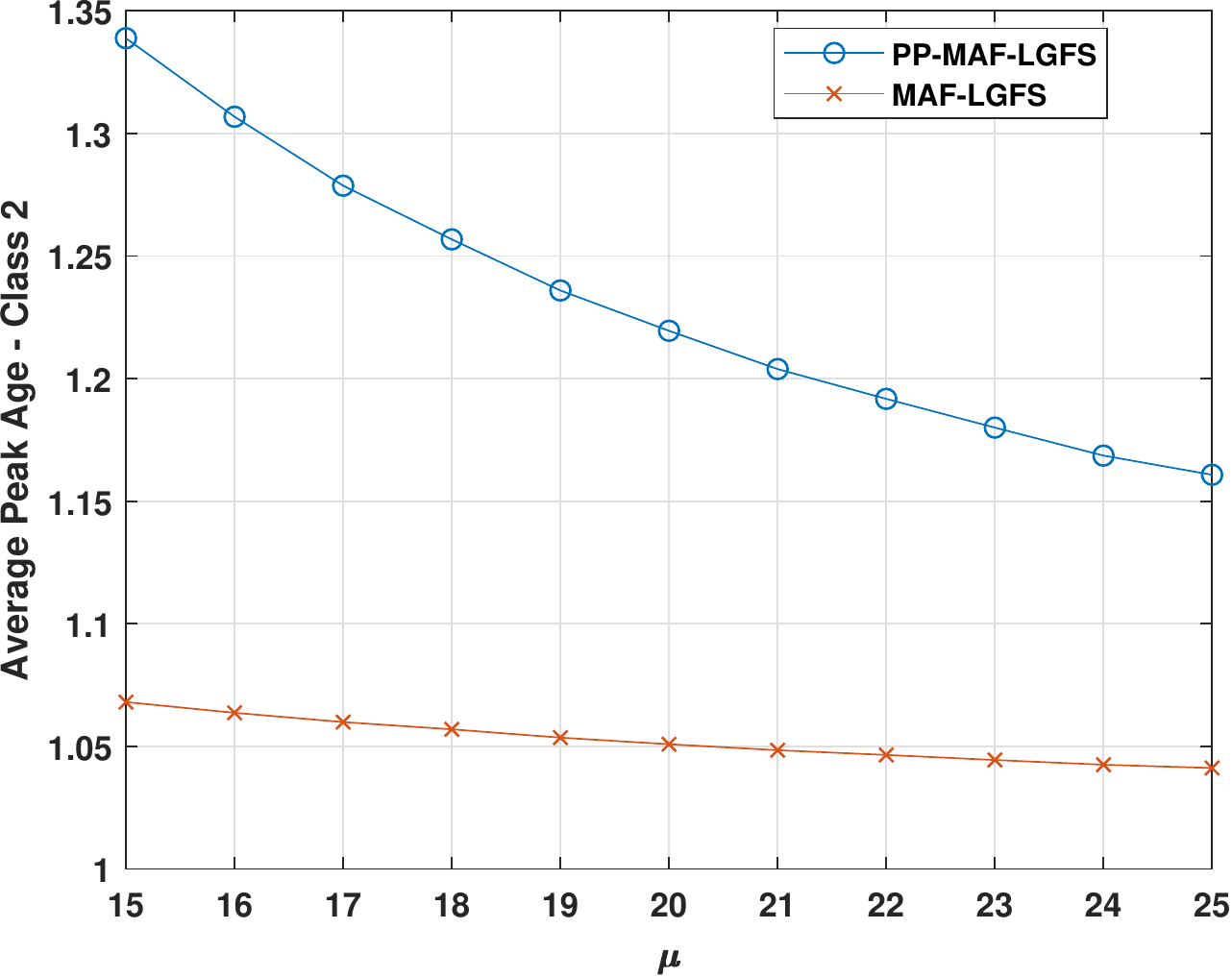}
  \caption{Average peak age of class $2$.}
\label{sim2}
\end{subfigure}%
\begin{subfigure}{0.3333\textwidth}
\centering
  \includegraphics[width=.99\linewidth]{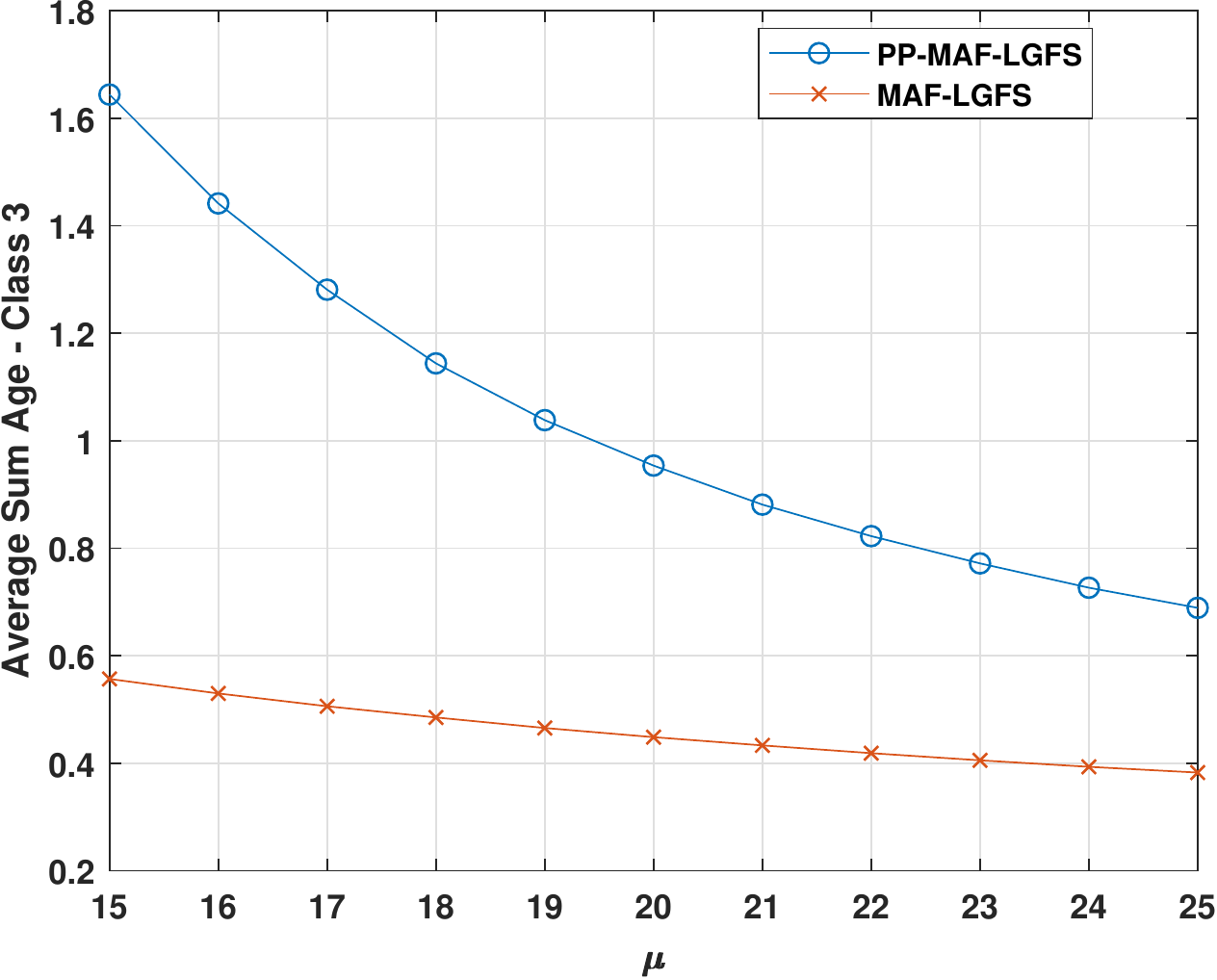}
  \caption{Average sum age of class $3$.}
\label{sim3}
\end{subfigure}%
\caption{Comparison between the two policies in function of the service rate $\mu$.}
\vspace{-20pt}
\label{simulationres}
\end{figure*}
We consider the case where the service time of each packet is exponentially distributed with service rate $\mu$. To address this multi-stream online scheduling problem, we first lay out the notion of informative packets.
\begin{definition}\emph{Informative and Non-informative Packets}:
Consider a packet of stream $(i,j)$ that is generated at time $S^{i,j}_n\leq t$. The packet is said to be informative (non-informative) at time $t$ if $t-S^{i,j}_n<\Delta^{i,j}(t)$ ($t-S^{i,j}_n\geq\Delta^{i,j}(t)$), i.e., the age of the packet is (not) smaller than $\Delta^{i,j}(t)$. 
%
%
%
%
%
%
%
%
%
%
%
%Consider a packet that arrives to the queue of stream $(i,j)$, and has been generated at time $S^{i,j}_n\leq t$. The packet is said to be informative (non-informative) at time $t$ 
%
%
%
%
%
%
%
%
%A packet belonging to a stream $(i,j)$ that has been generated at time $S^{i,j}_n$ and has arrived to the queue at time $A^{i,j}_n$ is said to be informative at time $t\geq A^{i,j}_n$ if
%\begin{equation}
%t-S^{i,j}_n<\Delta^{i,j}(t)
%\end{equation}
\end{definition}
Equipped with the above definition, we consider in the following several scheduling disciplines that are based on informative packets. 
\begin{definition}\emph{Preemptive Priority (PP) policy based on Informative Packets}: Among the streams with informative packets, the class of streams with the highest priority are served first. A packet in service is preempted upon the arrival of an informative packet of a higher priority stream; the preempted packet is stored back in the queue.
\end{definition}
\begin{definition}\textit{Maximum Age First (MAF) policy}: Among the streams from a priority class, the stream with the maximum age is served first, with ties broken arbitrarily.
\end{definition}
\begin{definition}\textit{Last-Generated, First-Served (LGFS) policy}: Among the informative packets from a stream, the last generated informative packet is served first, with ties broken arbitrarily.
\end{definition}
By combining the above three service disciplines, we propose a new scheduling policy called Preemptive Priority, Maximum Age First, Last-Generated, First-Served (\textbf{PP-MAF-LGFS}), which is defined as follows. 
%
%
%
%
%
%
%
%
%
%
%
%Our proposed scheduling policy, dubbed as Preemptive Priority, Maximum Age First, Last Generated, First Served (\textbf{PP-MAF-LGFS}), is work-conserving preemptive scheduling policy that is defined by combining the above selection mechanisms as follows.
\begin{definition}\textit{Preemptive Priority, Maximum Age First, Last-Generated, First-Served}: This policy is preemptive, work-conserving and obeys the following set of scheduling rules:
\begin{itemize}
\item If there exist informative packets, the system will serve an informative packet that is selected as follows
\begin{itemize}
\item among all streams with informative packets, pick the class of streams with the highest priority;
\item among the streams from the selected priority class, pick the stream with the maximum age, with ties broken arbitrarily;
\item among the informative packets from the selected stream, pick the last generated informative packet, with ties broken arbitrarily;
\end{itemize}
\item if there exists no informative packet, the system can serve any non-informative packet.
\end{itemize}
\label{definitionourpolicy}
\end{definition}
Note that our proposed policy does not drop non-informative packets as it was previously proposed in the literature (e.g., \cite{6875100}). Although these packets are not necessary for reducing the age, but in many applications, they may still be needed at the monitor (e.g., social updates). In the case of a single priority class (i.e., $I=1$), the proposed policy coincides with the Maximum Age First, Last-Generated, First-Served (\textbf{MAF-LGFS}) policy proposed in \cite{8406945}.\color{black}

By definition, our policy ensures that the service of high priority informative packets is not interrupted nor influenced by any lower priority packets. This grants crucial timely packets the best possible service by the facility. Note that informative packets play a key role in our policy. In particular, the preemptive priority discipline is a dynamic priority rule based on the existence of informative packets: If a stream from class $1$ has informative packets, the stream has the highest priority; otherwise, if the stream does not have any informative packets, the stream has the lowest priority, even lower than the streams in class $I$ that have informative packets. This non-trivial aspect of our policy ensures that low priority classes are provided with the best possible opportunity for transmission while not affecting the age of the high priority streams. On another note, our policy guarantees that the highest possible reduction in age from the selected priority class takes place at each packet delivery. These key observations are crucial and will be used to establish the lex-age-optimality of the PP-MAF-LGFS policy.  \color{black}
% Note that high priority classes are only picked based on the availability of informative packets. This allows low priority classes to be served when there are no informative packets for high priority classes. 
\begin{theorem}[Lex-age-optimality of PP-MAF-LGFS]
If (i) the
packet generation and arrival times are synchronized across streams within each class, and (ii) the packet service times are exponentially
distributed and i.i.d. across streams and time, then the policy \textup{PP-MAF-LGFS} is lex-age-optimal.
\label{theoremlexoptimal}
\end{theorem}
\begin{IEEEproof}
This theorem is proven using an inductive sample-path comparison. Specifically, we show by induction that the set of scheduling rules that the \textup{PP-MAF-LGFS} policy satisfies are sufficient and necessary for level $k$ lex-age-optimality for $k=1,\ldots,I$. Contrary to previous sample-path proofs in the literature (e.g., Theorem 1 in \cite{8406945}), showing these scheduling rules are sufficient for optimality is not enough in our case. In fact, at each induction step, a characterization of the exact behavior of each policy $\pi\in\Pi_{\text{lex-opt}}^{k}$ for $k=1,\ldots,I$ is required. This poses several technical difficulties, which we solve in our sample path proof by showing the necessity of the scheduling rules for level $k$ lex-age-optimality for $k=1,\ldots,I$. The details can be found in Appendix \ref{appendixpropositionmarkov}.
\end{IEEEproof}
Note that when each priority class has only one stream, the intra-class synchronization assumption is always satisfied and Theorem \ref{theoremlexoptimal} holds for \emph{arbitrarily given} packet generation and arrival times. This special case is of particular interest. 

To the best of our knowledge, this is the first lex-age-optimality results for multi-class status updates. Our results are strong as our optimality is established in terms of stochastic ordering of stochastic processes for all symmetric non-decreasing penalty functions, and for all non-decreasing age penalty functionals. What makes these results further interesting is that the priority classes can have different traffic patterns, age penalty functions, and age penalty functionals. As it was previously explained in Section II-C, this is of paramount importance as priority classes typically represent different applications, each with their own traffic arrivals and data timeliness requirements. For example, in a certain scenario, we can be interested in minimizing the peak max-age for class $1$, the time-average sum-age for class $2$, and the peak sum-age for class $3$. Theorem \ref{theoremlexoptimal} guarantees that our proposed policy achieves the required data timeliness goal for any of these cases, despite the differences in age penalty functions and functionals between the classes.
\section{Numerical Results}
We consider a vehicle in a V2X (Vehicle-To-Everything) network that sends packets to either nearby vehicles or roadside units (see \cite{5948952},\cite{VAHDATNEJAD201643} for two surveys). In the aforementioned surveys, a list of possible packets use cases are presented, each of which having different priorities in the network. We consider $3$ data categories in our simulations:
\begin{enumerate}
\item \textbf{Road Safety Data}: These are the data primarily employed to reduce the number of traffic accidents. These packets are generated periodically with a minimum frequency of $\SI{10}{\hertz} $. We assume in our settings that the packets' generation frequency is set to $\SI{10}{\hertz} $. This class of streams has the highest priority among all data types. We consider in our simulations that two streams belong to this class (e.g., the vehicle's position and speed).
\item \textbf{Traffic Management Data}: The goal of these data is to optimize the traffic stream and reduce the travel time in the network. We consider in our simulations that one stream belongs to this class (e.g., updates concerning the destination of the vehicle). The generation frequency of these packets is set to $\SI{1}{\hertz}$. The priority of this class is second to the road safety class.
\item \textbf{Convenience and Entertainment Data}: The data in this class are considered to be the least crucial as their aim is to provide entertainment and convenience solely for improving the quality of travel. We consider in our simulations that two streams belong to this class and we suppose that the generation frequency of their packets is $\SI{5}{\hertz}$.
\end{enumerate}
Based on the above, we can conclude that the arrival rate to our considered system is $\lambda_{tot}=31$ packets per second. The service facility of the vehicle is supposed to be constituted of $1$ server with the transmission times being i.i.d. across streams and time. Moreover, the transmission times are considered to be exponentially distributed with service rate $\mu$.

We compare our proposed policy to the preemptive MAF-LGFS\footnote{First-Come-First-Served (FCFS) policies are omitted from our simulations as they will always be outperformed by LGFS policies since queuing will lead to unnecessary staleness of the packets.} policy proposed in \cite{8406945}. The preemptive MAF-LGFS policy schedules the packet of the stream with the highest age, regardless of the class it belongs to. As for the age penalty function and functional for each class, we choose $p_{\text{exceed}-\alpha}$ and $\phi_{\text{avg}}$ as the age penalty function and functional for class $1$ respectively, where $\alpha$ is set to $\SI{250}{\ms}$. By doing so, we get
\begin{align}
&\mathbb{E}[\phi_{\text{avg}}(\{p_{\text{exceed}-\alpha}\circ\boldsymbol{\Delta}^{1}(t), t\geq0\})]\nonumber\\=&\frac{1}{2}\sum_{j=1}^{2}\frac{1}{T}\int_{0}^{T}\Pr(\Delta^{1,j}(t)>\alpha)dt,
%
%\frac{1}{T}\int_{0}^{T}p_{\text{exceed}-\alpha}\circ\boldsymbol{\Delta}^{1}(t)dt=\sum_{j=1}^{2}\frac{1}{2T}\int_{0}^{T}\mathbbm{1}_{\{\Delta^{1,j}(t)>\alpha\}}dt\xrightarrow[T\to\infty]{}\frac{1}{2}\sum_{j=1}^{2}
\end{align}
\color{black} where $\Pr(\Delta^{1,j}(t)>\alpha)$ is the probability of violation of the maximum tolerated age $\SI{250}{\ms}$ by stream $(1,j)$ at time $t$. The interest in this time-average age penalty function is that in vehicular networks, small age for the velocity and position data can be tolerated but, after a certain value, the performance of the system starts deteriorating due to this aging. For class $2$, we choose $p_{\text{sum}}$ and $\phi_{\text{peak}}$ as the age penalty function and functional, respectively. In other words, we are interested in minimizing the average peak-age of class $2$. Lastly, we choose $p_{sum}$ and $\phi_{\text{avg}}$ for class $3$.
% We choose $p_{max}$, previously reported in (\ref{functional2}), as the age penalty functional of class $1$. This is in accordance with our previous argument that, typically, the system performance related to safety data is affected by the maximum age of the delivered updates. For the remaining classes, we choose $p_{sum}$, previously reported in (\ref{functional1}), as their age penalty functional. 
\color{black} We iterate over a range of the service rate $\mu$ and we run the simulations for $\SI{e5}{\s}$. We report in Fig. \ref{simulationres} the simulations results that showcase the performance of each policy. We can conclude from these results the following: 
\begin{itemize}
\item As seen in Fig. \ref{sim1}, our proposed policy always outperforms the preemptive MAF-LGFS policy for class $1$ at any service rate. Specifically, the probability of the age threshold violation by the preemptive MAF-LGFS policy is $3$ times higher than the one achieved by our policy. This is a consequence of our proposed policy's goal as it gives priority to minimizing the time-average age penalty of class $1$ regardless of the other remaining classes. 
\item On the other hand, we can see in Fig. \ref{sim2}-\ref{sim3} that the preemptive MAF-LGFS policy outperforms our proposed policy for classes $2$ and $3$. In fact, in our policy, giving the priority to class $1$ leads to a penalty for the remainder of the classes. However, we recall that the probability of violation of the age threshold in class $1$ for our policy is $3$ times less than the preemptive MAF-LGFS. Accordingly, the penalty incurred by the remaining classes is justified. Moreover, we can see that as $\mu$ increases, the gap between the two curves in both figures shrinks. The reason behind this is that class $1$'s packets finish transmission much faster the higher $\mu$ is. Consequently, in our proposed policy, the server will be able to finish serving class $1$ fast enough that it can start serving the other classes before new packets for class $1$ arrive to the system. This reduces the incurred penalty by the low priority classes due to the presence of the high priority streams.
%\item The trend in Fig. \ref{sim4} is the combination of the above two points. In total, we can see that the total average age penalty of the system is lower for the preemptive MAF-LGFS policy. However, as the priority among the classes is not captured by the total average age, the gap between the two curves is justifiable.
\end{itemize}

The above results highlight the performance of our proposed lex-age-optimal policy, and provide a new direction on age analysis in multi-class scheduling scenarios.\color{black}
\section{Conclusion}
In this paper, we have introduced the notion of lex-age-optimality that captures both the age-optimality and the order of time-cruciality between the streams in a general multi-class priority-based scheduling scenario. To that end, we have proposed an online scheduling policy in a general multi-class, multi-stream scheduling scenario. Using a sample-path argument, we were able to prove the lex-age-optimality of the proposed policy in the single exponential server case for any symmetric non-decreasing penalty function, and for all non-decreasing age penalty functionals. Numerical results were then presented to highlight the performance of our proposed policy. 
\bibliographystyle{IEEEtran}
\bibliography{trialout}
\appendices
\section{Proof of Theorem \ref{theoremlexoptimal}}
\label{appendixpropositionmarkov}
To establish this theorem, we first provide a set of scheduling rules and prove by induction, and using a sample-path comparison, that they are necessary and sufficient for level $k$ lex-age-optimality for $k=1,\ldots,I$. Afterward, we show that the PP-MAF-LGFS policy satisfies these rules for all $k=1,\ldots,I$, and we can therefore conclude that it is lex-age-optimal. Before proceeding in this direction, we lay out some preliminaries on stochastic ordering that will be useful to our proof. \\
$\bullet$ \textbf{Preliminaries}: Let us consider two scheduling policies $P,\pi\in\Pi$. In general, for any class $i$, a direct comparison between two processes $\{p_t\circ\boldsymbol{\Delta}_{P}^{i}(t),t\geq0\}$ and $\{p_t\circ\boldsymbol{\Delta}_{\pi}^{i}(t),t\geq0\}$ to establish a stochastic ordering between the two is complex, as it involves comparing their probability distributions. To circumvent this difficulty, the following approach can be adopted:
\begin{itemize}
\item Define two policies $P_1,\pi_1\in\Pi$ on the same probability space such that $\{p_t\circ\boldsymbol{\Delta}_{P_1}^{i}(t),t\geq0\}$ and $\{p_t\circ\boldsymbol{\Delta}_{P}^{i}(t),t\geq0\}$ (respectively $\{p_t\circ\boldsymbol{\Delta}_{\pi_1}^{i}(t),t\geq0\}$ and $\{p_t\circ\boldsymbol{\Delta}_{\pi}^{i}(t),t\geq0\}$ ) have the same distribution.
\item Proceed with a direct comparison between $\{p_t\circ\boldsymbol{\Delta}_{P_1}^{i}(t),t\geq0\}$ and $\{p_t\circ\boldsymbol{\Delta}_{\pi_1}^{i}(t),t\geq0\}$.
\end{itemize}
This approach is called \emph{coupling} in the scheduling literature, and we will adopt it in our proof. To that end, and using the memoryless property of the exponential distribution, we can
obtain the following coupling lemma.  
\begin{lemma}[{\cite[Lemma~1]{8406945}} Stochastic Coupling]For any given $\mathcal{I}$, consider two \emph{work-conserving} policies $P,\pi\in\Pi_{wc}$. If the service times
are exponentially distributed and i.i.d. across streams and time, then the following holds:
\begin{enumerate}
\item There exists a \emph{work-conserving} policy $P_1$ such that $\{\boldsymbol{\Delta}_{P_1}(t),t\geq0\}$ and $\{\boldsymbol{\Delta}_{P}(t),t\geq0\}$ have the same distribution.
\item There exists a \emph{work-conserving} policy $\pi_1$ such that $\{\boldsymbol{\Delta}_{\pi_1}(t),t\geq0\}$ and $\{\boldsymbol{\Delta}_{\pi}(t),t\geq0\}$ have the same distribution.
\item $P_1$ and $\pi_1$ are defined on the same probability space and, if a packet is delivered in policy $\pi_1$ at time $t$, then with probability $1$, a packet is delivered in policy $P_1$ at time $t$.\color{black}
\end{enumerate}
\label{couplinglemma1}
\end{lemma}
\noindent Next, we present in the following proposition a set of scheduling rules for the first $k$ classes with $k\in\{1,\ldots,I\}$. We show that a policy $P$ is level $k$ lex-age-optimal if, and only if, these rules hold for the first $k$ classes. Note that, throughout this proof, we refer to classes $1$ till $k$ as the first $k$ classes. Before laying out the proposition, we define the notion of work-conserving policies for the informative packets of a class $k$.
\begin{definition}\textit{Work-conserving policies for the informative packets of a class $k$}:
A scheduling policy $P$ is said to be work-conserving for the informative packets of a class $k$ if the service facility is kept busy whenever there exist one or more informative packet in the queues of class $k$.
\end{definition}
\color{black}
\begin{proposition}[Lex-age-optimal Scheduling Rules]
If (i) the
packet generation and arrival times are synchronized across streams within each class, and (ii) the packet service times are exponentially
distributed and i.i.d. across streams and time, a scheduling policy $P$ is level $k$ lex-age-optimal for $k\in\{1,\ldots,I\}$ if, and only if, the following four \textbf{rules} are satisfied 
\begin{enumerate}
\item Policy $P$ is work-conserving for the informative packets of the first $k$ classes;
\item Among the streams with informative packets, $P$ serves the streams belonging to the first $k$ classes first. Among these classes with informative packets, the class of streams with the highest priority are preemptively served first;
\item Among the streams of each of the first $k$ classes with informative packets, the stream with the maximum age is served first, with ties broken arbitrarily;.
\item Among the informative packets from a stream of the first $k$ classes, the last generated informative packet is preemptively served first, with ties broken arbitrarily.
\end{enumerate}  
\end{proposition}
\begin{IEEEproof} 
We prove this proposition by induction. Specifically, we show in step $1$ that a policy is level $1$ lex-age-optimal if, and only if, Rules 1)-4) hold for class $1$. Then, by assuming that they are necessary and sufficient for level $k$ lex-age-optimality, we use this assumption to prove in step $2$ that these rules are sufficient and necessary for level $k+1$ lex-age-optimality.

\vspace{15pt}
\noindent$\bullet$ \textbf{Step 1}: We prove in this step that these rules for $k=1$ are sufficient and necessary for level $1$ lex-age-optimality.

\textbf{1) \emph{Sufficiency}}: Let us consider a work-conserving policy $P\in\Pi_{wc}$ that satisfies these rules for class $1$. We compare its performance to any \emph{work-conserving} policy $\pi\in\Pi_{wc}$. As both policies are work-conserving, we consider the two policies $P_1$ and $\pi_1$ that are defined on the same probability space and originate from Lemma \ref{couplinglemma1}. Next, we provide the following lemma that describes the evolution of the age vector of class $1$ upon a packet delivery by both $P_1$ and $\pi_1$.
\begin{lemma}[Packet Delivery] Suppose that a packet is delivered at time $t$ by both policies $\pi_1$ and $P_1$. The age vector changes at time $t$ from $\boldsymbol{\Delta}_{P_1}$ and $\boldsymbol{\Delta}_{\pi_1}$ to $\boldsymbol{\Delta}'_{P_1}$ and $\boldsymbol{\Delta}'_{\pi_1}$, respectively. If
\begin{equation}
\Delta^{1,[j]}_{P_1}\leq \Delta^{1,[j]}_{\pi_1}, \quad j=1,\ldots,J_1,
\label{conditionlema}
\end{equation}
then
\begin{equation}
(\Delta_{P_1}^{1,[j]})'\leq (\Delta^{1,[j]}_{\pi_1})', \quad j=1,\ldots,J_{1},
\label{desired15}
\end{equation}
\label{comparisondelivery}
where $\Delta^{1,[j]}_{P_1}$ and $\Delta^{1,[j]}_{\pi_1}$ refers to the $j$-th largest element of the age vector of class $1$ in policy $P_1$ and $\pi_1$, respectively. 
\end{lemma}
\begin{IEEEproof}
The proof can be found in Appendix \ref{proofoffirstlemma}.
\end{IEEEproof}
We can now proceed to prove that $P$ is level $1$ lex-age-optimal. To do so, we compare the age vector $\boldsymbol{\Delta}^{1}$ on a sample-path of the policies $P_1$ and $\pi_1$. We note that for any sample-path, $\boldsymbol{\Delta}_{P_1}(0^-)=\boldsymbol{\Delta}_{\pi_1}(0^-)$. To that end, We consider two cases:\\
Case $1$: When there are no packets deliveries by any of the policies, the age of each stream belonging to class $1$ increases at a unit rate.\\ 
Case $2$: When a packet is delivered by $\pi_1$, the evolution of the age vector of class $1$ is dictated by Lemma \ref{comparisondelivery}.
By induction over time, we obtain
\begin{equation}
\Delta^{1,[j]}_{P_1}(t)\leq \Delta^{1,[j]}_{\pi_1}(t), \quad j=1,\ldots,J_1, \quad t\geq0.
\label{inductione time}
\end{equation}
For any symmetric non-decreasing function $p_t$, and for $t\geq0$, it holds from 
(\ref{inductione time}) that
\begin{align}
&p_t\circ \boldsymbol{\Delta}^{1}_{P_1}(t)\nonumber\\
=&p_t(\Delta^{1,1}_{P_1}(t),\ldots,\Delta^{1,J_1}_{P_1}(t)) \nonumber\\
=&p_t(\Delta^{1,[1]}_{P_1}(t),\ldots,\Delta^{1,[J_1]}_{P_1}(t)) \nonumber\\
\leq &p_t(\Delta^{1,[1]}_{\pi_1}(t),\ldots,\Delta^{1,[J_1]}_{\pi_1}(t)) \nonumber\\
=&p_t(\Delta^{1,1}_{\pi_1}(t),\ldots,\Delta^{1,J_1}_{\pi_1}(t)) \nonumber\\
=&p_t\circ \boldsymbol{\Delta}^{1}_{\pi_1}(t).
\label{lastinequality}
\end{align}
By Lemma \ref{couplinglemma1}, the processes $\{\boldsymbol{\Delta}_{P_1}(t),t\geq0\}$ and $\{\boldsymbol{\Delta}_{P}(t),t\geq0\}$ (respectively the processes $\{\boldsymbol{\Delta}_{\pi_1}(t),t\geq0\}$ and $\{\boldsymbol{\Delta}_{\pi}(t),t\geq0\}$) have the same distribution. Accordingly, using (\ref{lastinequality}) and Theorem 6.B.30 in \cite{2007}, we can deduce that
\begin{align}
[\{p_t\circ\boldsymbol{\Delta}^{1}_{P}(t), t\geq0\}|\mathcal{I}]
\leq_{st}[\{p_t\circ\boldsymbol{\Delta}^{1}_{\pi}(t), t\geq0\}|\mathcal{I}],
\label{prooflevel1form}
\end{align}
for all $\mathcal{I}$, $p_t\in\mathcal{P}_{\text{sym}}$ and $\pi\in\Pi_{wc}$. The extension of (\ref{prooflevel1form}) to the case where $\pi$ is non-work-conserving is straightforward due to the exponential distribution of the service time and its independence across streams and time. In fact, due to the memoryless property offered by the exponential distribution, letting the server idle before a transmission will lead to unnecessary staleness of the available packets. This can be shown by a stochastic ordering argument but the details are omitted for the sake of space. Consequently, (\ref{prooflevel1form}) holds for any $\pi\in\Pi$ and, accordingly, $P$ is level $1$ lex-age-optimal.

\textbf{2) \emph{Necessity}}: In this part, we prove that every level $1$ lex-age-optimal policy satisfies these $4$ scheduling rules for class $1$. We do so by contradiction. Specifically, we consider a level $1$ lex-age-optimal policy $\pi\in\Pi_{\text{lex-opt}}^{1}$. We show that if $\pi$ violates any of these $4$ rules for class $1$, then it cannot be level $1$ lex-age-optimal.\\
- \textbf{Violation of Rule $\boldsymbol{1}$}: Let us consider that $\pi$ is not work-conserving for the informative packets of class $1$. Due to the memoryless property of the exponential distribution of the service time and its independence across streams and time, letting the server idle before a transmission will lead to unnecessary staleness of the available informative packets. This can be shown by a stochastic ordering argument but the details are omitted for the sake of space. Accordingly, $\pi$ cannot be level $1$ lex-age-optimal.\\
- \textbf{Violation of Rule $\boldsymbol{2-4}$}: As shown in the proof of necessity of Rule $1$, we can affirm that $\pi$ has to be work-conserving for the informative packets of class $1$. Note that when there are no informative packets for class $1$ in the system, the performance of class $1$'s streams is not affected by the scheduling rules adopted. Accordingly, and without loss of generality, let us consider that $\pi$ is \emph{work-conserving}. In other words, we have $\pi\in\Pi_{wc}\cap\Pi_{\text{lex-opt}}^1$. By Definition \ref{definitionlexoptimality} and (\ref{newdefinitionlevel1}), we have 
\begin{align}
&\mathbb{E}[\phi(\{p_t\circ\boldsymbol{\Delta}^{1}_{\pi}(t), t\geq0\})|\mathcal{I}]\nonumber\\=&\min_{\pi'\in \Pi}\mathbb{E}[\phi(\{p_t\circ\boldsymbol{\Delta}^{1}_{\pi'}(t), t\geq0\})|\mathcal{I}],
\label{mashimashi}
\end{align}
for all $\mathcal{I}$, $p_t\in\mathcal{P}_{\text{sym}}$ and non-decreasing functional $\phi:\mathbb{V}\mapsto \mathbb{R}$,
provided that the expectations in (\ref{mashimashi}) exist. We show by contradiction that if $\pi$ violates any of the rules $2-4$ for class $1$, then there exists a policy $P$, a symmetrical non-decreasing penalty function $p'$, and a non-decreasing functional $\phi_1$ such that
\begin{align}
&\mathbb{E}[\phi_1(\{p'\circ\boldsymbol{\Delta}^{1}_{P}(t), t\geq0\})|\mathcal{I}]\nonumber\\<&\mathbb{E}[\phi_1(\{p'\circ\boldsymbol{\Delta}^{1}_{\pi}(t), t\geq0\})|\mathcal{I}].
\label{contradictionwhatweneed}
\end{align}
To that end, let us consider a work-conserving policy $P$ that satisfies these $4$ rules for class $1$. Note that $P$ and $\pi$ are both work-conserving. Accordingly, we consider the two coupled policies $P_1$ and $\pi_1$ that are defined on the same probability space and originate from Lemma \ref{couplinglemma1}. From the sufficiency proof, (\ref{inductione time}) holds for our case. In other words,  
\begin{equation}
\Delta^{1,[j]}_{P_1}(t)\leq \Delta^{1,[j]}_{\pi_1}(t), \quad j=1,\ldots,J_1, \quad t\geq0.
\label{inductionetimenewww}
\end{equation}
Accordingly, for any symmetrical non-decreasing function $p_t\in\mathcal{P}_{\text{sym}}$, and for $t\geq0$
\begin{equation}
p_t\circ \boldsymbol{\Delta}^{1}_{P_1}(t)\leq p_t\circ \boldsymbol{\Delta}^{1}_{\pi_1}(t).
\label{newlemma1}
\end{equation}
Next, let us consider a delivery time $t_s$ such that (i) the age of streams of class $1$ are not all equal to one another\footnote{We avoid this scenario since, in the case where all streams have the same age, all streams of class $1$ are considered to have the highest age.}, and (ii) there exist informative packets for $l_1>0$ and $l_2>0$ streams of class $1$ in the system just before $t_s$ for policy $\pi_1$ and $P_1$, respectively. As $P_1$ follows the $4$ rules of the proposition for class $1$, we have $l_2\leq l_1$. We recall that, according to Lemma \ref{couplinglemma1}, if a packet is delivered in policy $\pi_1$ at time $t$, then with probability $1$, a packet is delivered in policy $P_1$ at time $t$. Hence, we describe the evolution of the age vector of class $1$ upon a packet delivery by both policies $\pi_1$ and $P_1$ at time $t_s$.
\begin{lemma}[Packet Delivery]
Suppose that a packet is delivered
at time $t_s$ by both policies $\pi_1$ and $P_1$. The age vector changes at time $t_s$ from $\boldsymbol{\Delta}_{P_1}$ and $\boldsymbol{\Delta}_{\pi_1}$ to $\boldsymbol{\Delta}'_{P_1}$ and $\boldsymbol{\Delta}'_{\pi_1}$, respectively. If $\pi_1$ breaks any of the scheduling rules $2-4$ for class $1$ at time $t_s$, then there exists a stream $j$ of class $1$ such that
\begin{equation}
(\Delta_{P_1}^{1,[j]})'<(\Delta^{1,[j]}_{\pi_1})'.
\label{desirednew}
\end{equation}
\label{comparisondeliverynew}
\end{lemma}
\begin{IEEEproof}
The proof can be found in Appendix \ref{prooflemmatenye}.
\end{IEEEproof}
\noindent Next, to prove (\ref{contradictionwhatweneed}), let us consider the symmetrical non-decreasing penalty function $p'=p_{\text{sum}}\in\mathcal{P}_{\text{sym}}$ and the non-decreasing age penalty functional $\phi_1=\phi_{avg}$. By taking Lemma \ref{comparisondeliverynew} into account, along with
(\ref{inductionetimenewww}), and the fact that the service rate $\mu$ is finite, we can affirm that there exists a time interval $\mathscr{T}\subseteq[0,\infty)$ such that
\begin{equation}
p'\circ\boldsymbol{\Delta}^{1}_{P_1}(t)<p'\circ\boldsymbol{\Delta}^{1}_{\pi_1}(t)\quad \forall t\in\mathscr{T}.
\label{ekherstrict}
\end{equation}
By Lemma \ref{couplinglemma1}, we have that the processes $\{\boldsymbol{\Delta}_{P_1}(t),t\geq0\}$ and $\{\boldsymbol{\Delta}_{P}(t),t\geq0\}$ (respectively the processes $\{\boldsymbol{\Delta}_{\pi_1}(t),t\geq0\}$ and $\{\boldsymbol{\Delta}_{\pi}(t),t\geq0\}$) have the same distribution. By taking this into account, and by using (\ref{newlemma1}) and (\ref{ekherstrict}), we obtain:
\begin{equation}
\mathbb{E}[\phi_1(\{p'\circ\boldsymbol{\Delta}^{1}_{P}(t), t\geq0\}|\mathcal{I})]<\mathbb{E}[\phi_1(\{p'\circ\boldsymbol{\Delta}^{1}_{\pi}(t), t\geq0\}|\mathcal{I})]
\end{equation}
Therefore, $\pi$ is not level $1$ lex-age-optimal if it breaks any of the $4$ scheduling rules of the proposition for class $1$.
\\ \noindent This concludes our proof that this set of rules for class $1$ are sufficient and necessary to have level $1$ lex-age-optimality.

 \vspace{15pt}
\noindent$\bullet$\textbf{{Step $\boldsymbol{2}$}}: Next, we will prove the induction step: Assume that this set of rules for the first $k$ classes are necessary and sufficient for level $k$ lex-age-optimality. In other words, every policy $\pi\in\Pi_{\text{lex-opt}}^{k}$ follows these scheduling rules for the first $k$ classes. Our goal is to use this assumption to prove that a policy $P$ is level $k+1$ lex-age-optimal if, and only if, it follows these rules for the first $k+1$ classes.

%Towards this goal, we divide our proof to two parts 1) sufficiency, and 2) necessity.\\
\textbf{1) \emph{Sufficiency}}: Let us consider a work-conserving policy $P$ that satisfies the depicted set of rules for the first $k+1$ classes. We compare its performance to any \emph{work-conserving} policy $\pi\in\Pi_{wc}\cap\Pi_{\text{lex-opt}}^{k}$. As both policies are work-conserving, we consider the two policies $P_1$ and $\pi_1$ that are defined on the same probability space and originate from Lemma \ref{couplinglemma1}. Next, we provide the following Lemma that describes the evolution of the age vector of classes $i=1,\ldots,k+1$ upon a packet delivery by both $\pi_1$ and $P_1$.
\begin{lemma}[Packet Delivery] Suppose that a packet is delivered at time $t$ by both policies $\pi_1$ and $P_1$. The age vector changes at time $t$ from $\boldsymbol{\Delta}_{P_1}$ and $\boldsymbol{\Delta}_{\pi_1}$ to $\boldsymbol{\Delta}'_{P_1}$ and $\boldsymbol{\Delta}'_{\pi_1}$, respectively. If
\begin{align}
\Delta^{i,[j]}_{P_1}&=\Delta^{i,[j]}_{\pi_1}, \quad i=1,\ldots,k, \:\:j=1,\ldots,J_i,
\label{conditionlemafork1}\\\Delta^{k+1,[j]}_{P_1}&\leq \Delta^{k+1,[j]}_{\pi_1}, \quad j=1,\ldots,J_{k+1},
\label{conditionlemafork2}
\end{align}
then
\begin{align}
(\Delta_{P_1}^{i,[j]})'&=(\Delta^{i,[j]}_{\pi_1})', \quad i=1,\ldots,k, \:\:j=1,\ldots,J_i,\label{desired15fork1}\\
(\Delta_{P_1}^{k+1,[j]})'&\leq (\Delta^{k+1,[j]}_{\pi_1})', \quad j=1,\ldots,J_{k+1}.
\label{desired15fork2}
\end{align}
\label{comparisondeliveryfork}
\end{lemma}
\begin{IEEEproof}
The proof can be found in Appendix \ref{prooflemmatelte}.
\end{IEEEproof}
We can now show that $P$ is level $k+1$ lex-age-optimal. To do so, we compare the age vector $\boldsymbol{\Delta}^{i}$ for $i=1,\ldots,k+1$ on a sample-path of the policies $P_1$ and $\pi_1$. We note that for any sample-path, $\boldsymbol{\Delta}_{P_1}(0^-)=\boldsymbol{\Delta}_{\pi_1}(0^-)$. To that end, we consider two cases:\\
Case $1$: When there are no packets deliveries by any of the policies, the age of each stream of the first $k+1$ classes increases at a unit rate.\\ 
Case $2$: When a packet is delivered by $\pi_1$, the evolution of the age vector of the first $k+1$ classes is dictated by Lemma \ref{comparisondeliveryfork}.
By induction over time, we obtain for all $t\geq0$:
\begin{align}
\Delta^{i,[j]}_{P_1}(t)&= \Delta^{i,[j]}_{\pi_1}(t), \quad i=1,\ldots,k, \:\: j=1,\ldots,J_i, \label{inductione timekequal}\\
\Delta^{k+1,[j]}_{P_1}(t)&\leq \Delta^{k+1,[j]}_{\pi_1}(t), \quad j=1,\ldots,J_{k+1}. 
\label{inductione timeklessorequal}
\end{align}
For any symmetric non-decreasing function $p_t$, and for $t\geq0$, it holds from 
(\ref{inductione timekequal}) and (\ref{inductione timeklessorequal}) 
\begin{align}
p_t\circ\boldsymbol{\Delta}^{i}_{P_{1}}(t)&=p_t\circ\boldsymbol{\Delta}^{i}_{\pi_1}(t), \quad i=1,\ldots,k, \quad t\geq0,
\label{ptequalityk}\\
p_t\circ\boldsymbol{\Delta}^{k+1}_{P_{1}}(t)&\leq p_t\circ\boldsymbol{\Delta}^{k+1}_{\pi_1}(t), \quad t\geq0.
\label{ptinequalityk}
\end{align}
By Lemma \ref{couplinglemma1}, the processes $\{\boldsymbol{\Delta}_{P_1}(t),t\geq0\}$ and $\{\boldsymbol{\Delta}_{P}(t),t\geq0\}$ (respectively the processes $\{\boldsymbol{\Delta}_{\pi_1}(t),t\geq0\}$ and $\{\boldsymbol{\Delta}_{\pi}(t),t\geq0\}$) have the same distribution. Accordingly, using (\ref{ptequalityk})-(\ref{ptinequalityk}) and Theorem 6.B.30 in \cite{2007}, we can deduce that
\begin{align}
&[\{p_t\circ\boldsymbol{\Delta}^{i}_{P}(t), t\geq0\}|\mathcal{I}]\nonumber\\
=_{st}&[\{p_t\circ\boldsymbol{\Delta}^{i}_{\pi}(t), t\geq0\}|\mathcal{I}], \quad i=1,\ldots,k,
\label{prooflevelkform}\\
\text{and\quad\quad\quad\quad\:\:}\nonumber\\
&[\{p_t\circ\boldsymbol{\Delta}^{k+1}_{P}(t), t\geq0\}|\mathcal{I}]\nonumber\\
\leq_{st}&[\{p_t\circ\boldsymbol{\Delta}^{k+1}_{\pi}(t), t\geq0\}|\mathcal{I}], 
\label{prooflevelkforminequality}
\end{align}
for all $\mathcal{I}$, $p_t\in\mathcal{P}_{\text{sym}}$ and $\pi\in\Pi_{wc}\cap\Pi_{\text{lex-opt}}^{k}$. The extension of (\ref{prooflevelkform})-(\ref{prooflevelkforminequality}) to the case where $\pi\in\Pi_{\text{lex-opt}}^{k}$ but is not necessarily work-conserving is straightforward due to the exponential distribution of the service time and its independence across streams and time. As it was previously explained, due to the memoryless property offered by the exponential distribution, letting the server to idle before a transmission will lead to unnecessary staleness of the packets. This can be shown by a stochastic ordering argument but the details are omitted for the sake of space. Consequently, (\ref{prooflevelkform})-(\ref{prooflevelkforminequality}) hold for any $\pi\in\Pi_{\text{lex-opt}}^{k}$ and, therefore, $P$ is level $k+1$ lex-age-optimal.
%
%
%The extension of (\ref{prooflevelkform})-(\ref{prooflevelkforminequality}) to the case where $\pi$ is not necessarily work-conserving for the informative packets of class $k+1$ is straightforward due to the exponential distribution of the service time and its independence across streams and time. As it was previously explained, due to the memoryless property offered by the exponential distribution, letting the server to idle before a transmission will lead to unnecessary staleness of the available informative packets of class $k+1$. This can be shown by a stochastic ordering argument but the details are omitted for the sake of space. Note that (\ref{prooflevelkform})-(\ref{prooflevelkforminequality}) remain valid regardless of what $\pi$ does when there are no informative packets for the first $k+1$ classes. Consequently, (\ref{prooflevelkform})-(\ref{prooflevelkforminequality}) hold for any $\pi\in\Pi_{\text{lex-opt}}^{k}$ and, therefore, $P$ is level $k+1$ lex-age-optimal.
%

\textbf{2) \emph{Necessity}}: In this part, we leverage our inductive assumption for level $k$ lex-age-optimality and prove that every level $k+1$ lex-age-optimal policy follow these $4$ scheduling rules for the first $k+1$ classes. We prove this by contradiction. Specifically, let us consider a level $k$ lex-age-optimal policy $\pi\in\Pi_{\text{lex-opt}}^{k}$. We know by our inductive assumption that $\pi$ has to follow this set of rules for the first $k$ classes. We show that if $\pi$ violates any of the $4$ rules for class $k+1$, then it cannot be level $k+1$ lex-age-optimal.\\
- \textbf{Violation of Rule $\boldsymbol{1}$}: Let us consider that $\pi$ is not work-conserving for the informative packets of class $k+1$. Due to the memoryless property of the exponential distribution of the service time and its independence across streams and time, letting the server idle before a transmission will lead to unnecessary staleness of the available packets. This can be shown by a stochastic ordering argument but the details are omitted for the sake of space. Accordingly, $\pi$ cannot be level $k+1$ lex-age-optimal.\\
- \textbf{Violation of Rule $\boldsymbol{2-4}$}: The proof follows the same line of work done in the necessity proof of Step $1$. Specifically, and as it was previously explained, we can consider that $\pi\in\Pi_{wc}\cap\Pi_{\text{lex-opt}}^k$. Next, we consider a work-conserving policy $P$ that satisfies the $4$ scheduling rules for the first $k+1$ classes. Note that $P$ and $\pi$ are both work-conserving. Accordingly, we consider the two coupled policies $P_1$ and $\pi_1$ that are defined on the same probability space and originate from Lemma \ref{couplinglemma1}. From the sufficiency proof for level $k+1$ lex-age-optimality, we have that for all $t\geq0$:
\begin{align}
\Delta^{i,[j]}_{P_1}(t)&= \Delta^{i,[j]}_{\pi_1}(t), \quad i=1,\ldots,k, \:\: j=1,\ldots,J_i, \label{inductione timekequalnnnnn}\\
\Delta^{k+1,[j]}_{P_1}(t)&\leq \Delta^{k+1,[j]}_{\pi_1}(t), \quad j=1,\ldots,J_{k+1}. 
\label{inductione timeklessorequalnnnnn}
\end{align}
Accordingly, for any symmetric non-decreasing function $p_t$:
\begin{align}
p_t\circ\boldsymbol{\Delta}^{i}_{P_{1}}(t)&=p_t\circ\boldsymbol{\Delta}^{i}_{\pi_1}(t), \quad i=1,\ldots,k \quad t\geq0,
\label{sufffptequalityk}\\
p_t\circ\boldsymbol{\Delta}^{k+1}_{P_{1}}(t)&\leq p_t\circ\boldsymbol{\Delta}^{k+1}_{\pi_1}(t), \quad t\geq0.
\label{sufffptinequalityk}
\end{align}
Next, as per our inductive assumption, we have that $\pi_1$ and $P_1$ follow the same scheduling discipline for the first $k$ classes. Accordingly, the streams of the first $k$ classes will have no informative updates at the same time in both policies $\pi_1$ and $P_1$. This allows us to consider a delivery time $t_s$ such that (i) there are no informative packets for the first $k$ classes, (ii) the age of streams of class $k+1$ are not all equal to one another, and (iii) there exist informative packets for $l_1>0$ and $l_2>0$ streams of class $k+1$ in the system just before $t_s$ for policy $\pi_1$ and $P_1$, respectively. As $P_1$ follows the $4$ scheduling rules of the proposition for the first $k+1$ classes, we have $l_2\leq l_1$. By proceeding similarly to Lemma \ref{comparisondeliverynew}, we can show that if $\pi_1$ breaks any of the scheduling rules $2-4$ for class $k+1$ at time $t_s$, then there exists a stream $j$ of class $k+1$ such that
\begin{equation}
\Delta_{P_1}^{k+1,[j]}(t_s^+)<\Delta^{k+1,[j]}_{\pi_1}(t_s^+).
\label{desirednewnewnew}
\end{equation}
Afterward, we consider the symmetric non-decreasing penalty function $p'=p_{\text{sum}}\in\mathcal{P}_{\text{sym}}$ and the non-decreasing age penalty functional $\phi_1=\phi_{avg}$. By taking (\ref{desirednewnewnew}) into account, along with
(\ref{inductione timekequalnnnnn})-(\ref{inductione timeklessorequalnnnnn}), and the fact that the service rate $\mu$ is finite, we can affirm that there exists a time interval $\mathscr{T}\subseteq[0,\infty)$ such that
\begin{equation}
p'\circ\boldsymbol{\Delta}^{k+1}_{P_1}(t)<p'\circ\boldsymbol{\Delta}^{k+1}_{\pi_1}(t),\quad \forall t\in\mathscr{T}.
\label{ekherstrictnew}
\end{equation}
By Lemma \ref{couplinglemma1}, we have that the processes $\{\boldsymbol{\Delta}_{P_1}(t),t\geq0\}$ and $\{\boldsymbol{\Delta}_{P}(t),t\geq0\}$ (respectively the processes $\{\boldsymbol{\Delta}_{\pi_1}(t),t\geq0\}$ and $\{\boldsymbol{\Delta}_{\pi}(t),t\geq0\}$) have the same distribution. By taking this into consideration, and by using (\ref{sufffptequalityk}), (\ref{sufffptinequalityk}), and (\ref{ekherstrictnew}), we obtain:
%\begin{align}
%&[\{p_t\circ\boldsymbol{\Delta}^{i}_{P}(t), t\geq0\}|\mathcal{I}]\nonumber\\
%=_{st}&[\{p_t\circ\boldsymbol{\Delta}^{i}_{\pi}(t), t\geq0\}|\mathcal{I}], \quad i=1,\ldots,k,
%\label{prooflevelkformekhershi}
%\end{align}
%and
%\begin{align}
%&\mathbb{E}[\phi_1(\{p'\circ\boldsymbol{\Delta}^{k+1}_{P}(t), t\geq0\}|\mathcal{I})]\nonumber\\<&\mathbb{E}[\phi_1(\{p'\circ\boldsymbol{\Delta}^{k+1}_{\pi}(t), t\geq0\}|\mathcal{I})].
%\end{align}
\begin{align}
&[\{p_t\circ\boldsymbol{\Delta}^{i}_{P}(t), t\geq0\}|\mathcal{I}]\nonumber\\
=_{st}&[\{p_t\circ\boldsymbol{\Delta}^{i}_{\pi}(t), t\geq0\}|\mathcal{I}], \quad i=1,\ldots,k,
\label{prooflevelkformekhershi}\\
\text{and\quad\quad\quad\quad\:\:}\nonumber\\
&\mathbb{E}[\phi_1(\{p'\circ\boldsymbol{\Delta}^{k+1}_{P}(t), t\geq0\}|\mathcal{I})]\nonumber\\<&\mathbb{E}[\phi_1(\{p'\circ\boldsymbol{\Delta}^{k+1}_{\pi}(t), t\geq0\}|\mathcal{I})].
\end{align}
Therefore, $\pi$ is not level $k+1$ lex-age-optimal if it breaks any of the $4$ scheduling rules of the proposition for class $k+1$.
\end{IEEEproof}
By Definition \ref{definitionourpolicy}, the PP-MAF-LGFS policy is the only policy that satisfies the scheduling rules depicted in this proposition for the first $k$ classes simultaneously for any $k=1,\ldots,I$. Accordingly, the PP-MAF-LGFS policy is lex-age-optimal, which concludes the proof of the theorem.
\section{Proof of Lemma \protect\ref{comparisondelivery}}
\label{proofoffirstlemma}
Let us denote by $W^1_{j}(t)=\max\{S^{1,j}_n:A^{1,j}_n\leq t\}$ the time-stamp of the freshest packet that has arrived to the queue of stream $j$ of class $1$ at time $t$. Since the generation/arrival sequences are synchronized across streams within each class, there exists a $W^1(t)$ such that $W^1_{j}(t)=W^1(t)$ for $j=1,\ldots,J_{1}$. We distinguish between three cases that can happen at time $t$. The proof of Case $3$ is adopted from the proof of Lemma 2 of \cite{8406945}. For the sake of completeness, we provide a proof of all $3$ cases. 
\subsubsection{Case 1} There was no transmission of packets for class $1$ by policy $P_1$, or a non-informative packet of class $1$ has just finished transmission. In other words, prior to time $t$, policy $P_1$ has already finished the transmission of all class $1$'s informative packets. \color{black}To that end:
\begin{equation}
(\Delta^{1,[j]}_{P_1})'=\Delta^{1,[j]}_{P_1}=t-W^1(t),\quad j=1,\ldots,J_1.
\end{equation}
On the other hand, in policy $\pi_1$, the delivered packet can be any packet from any information stream. Consequently, we can conclude:
\begin{equation}
\Delta^{1,[j]}_{\pi_1}\geq (\Delta^{1,[j]}_{\pi_1})'\geq t-W^1(t),\quad j=1,\ldots,J_1.
\end{equation}
Therefore, (\ref{desired15}) holds for this case.
\subsubsection{Case 2} An informative \color{black}packet belonging to a stream of class $1$ finishes transmission by policy $P_1$ at time $t$. On the other hand, policy $\pi_1$ delivers a non-informative packet of class $1$ or a packet belonging to one of the $I-1$ remaining classes at time $t$. Consequently, $(\boldsymbol{\Delta}^1_{\pi_1})'=\boldsymbol{\Delta}^1_{\pi_1}$ and (\ref{desired15}) holds trivially in this scenario.
\subsubsection{Case 3} An informative \color{black}packet belonging to a stream of class $1$ finishes transmission by both policies $P_1$ and $\pi_1$ at time $t$. By definition, the following always holds:
\begin{equation}
\Delta^{1,j}_{P_1}\geq(\Delta^{1,j}_{P_1})'\geq t-W^1(t), \quad j=1,\ldots,J_1,
\end{equation}
\begin{equation}
\Delta^{1,j}_{\pi_1}\geq (\Delta^{1,j}_{\pi_1})'\geq t-W^1(t),\quad j=1,\ldots,J_1.
\label{holdsalways}
\end{equation}
We recall that $P_1$ schedules the stream of class $1$ with the highest age. Consequently, the stream of class $1$ having the age $\Delta^{1,[1]}_{P_1}$ is the one that finishes transmission at time $t$ by $P_1$. Since the transmitted packet has $W^1(t)$ as time-stamp, the age of this stream becomes the smallest among the streams of class $1$. To that end, 
\begin{equation}
(\Delta^{1,[J_1]}_{P_1})'=t-W^1(t).
\label{sarazghar}
\end{equation}
As there is only one server, the age of the remaining $J_1-1$ streams of class $1$ stay the same. By taking this into account, along with (\ref{sarazghar}), we get: 
\begin{equation}
(\Delta^{1,[j]}_{P_1})'=\Delta^{1,[j+1]}_{P_1}, \quad j=1,\ldots,J_1-1.
\label{nafslshi}
\end{equation} 
On the other hand, since the packet delivered by $\pi_1$ can belong to any stream of class $1$, the following always holds:
\begin{equation}
(\Delta^{1,[j]}_{\pi_1})'\geq\Delta^{1,[j+1]}_{\pi_1}, \quad j=1,\ldots,J_1-1.
\label{piiteration}
\end{equation}
Combining (\ref{conditionlema}), (\ref{nafslshi}) and (\ref{piiteration}), we obtain:
\begin{equation}
(\Delta^{1,[j]}_{\pi_1})'\geq\Delta^{1,[j+1]}_{\pi_1}\geq\Delta^{1,[j+1]}_{P_1}=(\Delta^{1,[j]}_{P_1})', \:\: j=1,\ldots,J_1-1.
\end{equation}
Also, using (\ref{holdsalways}) and (\ref{sarazghar}), we can deduce that $(\Delta^{1,[J_i]}_{\pi_1})'\geq t-W^1(t)=(\Delta^{1,[J_i]}_{P_1})'$ which concludes the proof.
\section{Proof of Lemma \protect\ref{comparisondeliverynew}}
\label{prooflemmatenye}
To prove this lemma, we recall that (\ref{inductionetimenewww}) always holds from our sufficiency results on $P$. Next, we distinguish between $3$ cases.\\
Case $1$: Suppose that $\pi_1$ breaks Rule $2$ and delivers at time $t_s$ a packet that does not belong to class $1$. 
%In this case, the sorted age vector $[\boldsymbol{\Delta}^{1}_{\pi_1}]$ keeps the same order after the packet delivery at time $t_s$. 
We know that $P_1$ will deliver at time $t_s$ an informative packet for one of the $l_2$ streams belonging to class $1$. Consequently, (\ref{desirednew}) holds trivially in this case.\\
Case $2$: Suppose that $\pi_1$ delivers a packet from class $1$. However, at time $t_s$, $\pi_1$ breaks Rule $3$ for class $1$ and delivers a packet that does not belong to the stream of class $1$ with the highest age. To tackle this case, we define the rank of a stream within a class. 
\begin{definition}\textit{Rank of a stream}:
The rank of a stream $(i,j)$ within the class $i$ is defined as its position in the ordered age vector $[\boldsymbol{\Delta}^{i}]$. In other words, if stream $(i,j)$ has a rank $1\leq r\leq J_i$, then:
\begin{itemize}
\item There exist $J_i-r$ streams in the same class having an age that is smaller or equal to $\Delta^{i,j}$.
\item There exist $r-1$ streams in the same class having an age that is larger or equal to $\Delta^{i,j}$.
\end{itemize}
\end{definition}
\noindent We know that $P_1$ delivers the freshest packet from the stream of class $1$ with the highest age at time $t_s$ (i.e., the stream with rank $1$). Therefore, after delivery, the served stream will have the smallest age among all streams of class $1$. Moreover, the age of the remaining $J_1-1$ streams of class $1$ is not altered at the delivery time. Accordingly, these $J_1-1$ streams gain a single rank in the sorted age vector  $[\boldsymbol{\Delta}^{1}_{P_1}]$. On the other hand, let us suppose that the served stream by $\pi_1$ has a rank $r>1$ in the sorted age vector $[\boldsymbol{\Delta}^{1}_{\pi_1}]$. After being served, this stream will have a rank $r'\leq r$.  Consequently, $r'-r$ streams will gain a rank at time $t_s$ and the rank of all the remaining streams stays the same. Therefore, we can assert that (\ref{desirednew}) holds. We provide in the following an example to showcase this. Suppose that the ordered age vector of class $1$ just before $t_s$ is:
\begin{equation*}
[\boldsymbol{\Delta}^{1}_{\pi_1}](t_s^{-})=(10,9,8,1),
\end{equation*}
\begin{equation}
[\boldsymbol{\Delta}^{1}_{P_1}](t_s^{-})=(10,9,8,1).
\end{equation}
Suppose that the age of the available informative packets of class $1$ is equal to $1$ at time $t_s$. If we consider that $\pi_1$ delivers a packet from stream $(1,[3])$, and knowing that $P_1$ will deliver a packet from stream $(1,[1])$, we get:
\begin{equation*}
[\boldsymbol{\Delta}^{1}_{\pi'_1}](t_s^{+})=(10,9,1,1)
\end{equation*}
\begin{equation}
[\boldsymbol{\Delta}^{1}_{\pi_1}](t_s^{+})=(9,8,1,1)
\end{equation}
Accordingly, we can easily see that $j=1$ or $j=2$.\\
Case $3$: Suppose that $\pi_1$ delivers a packet from the stream of class $1$ with the highest age at time $t_s$. However, suppose that $\pi_1$ breaks Rule $4$ for class $1$ and does not deliver the freshest available informative packet. Accordingly, at time $t_s$, the served stream by $P_1$ will have a strictly smaller age when compared to the stream served by $\pi_1$. Consequently, (\ref{desirednew}) holds.
\section{Proof of Lemma \protect\ref{comparisondeliveryfork}}
\label{prooflemmatelte}
We proceed with our proof by distinguishing between two possible scenarios at time $t$:
\begin{itemize}
\item \emph{The served packet by $\pi_1$ is an informative packet belonging to any of the first $k$ classes}: We recall that, as per our inductive assumption till level $k$, policy $\pi_1$ and $P_1$ follow the same set of scheduling rules for the first $k$ classes. Accordingly, when an informative packet from one of these classes is delivered by $\pi_1$, the same packet (or an informative packet of another stream of the same class that has the same age) is delivered by $P_1$. Consequently, we can affirm the validity of (\ref{desired15fork1}). Moreover, as the age vector of class $k+1$ remains unchanged for both policies in this case, (\ref{desired15fork2}) holds naturally.
\item \emph{The served packet by $\pi_1$ is not an informative packet belonging to the $k$ first classes}: As $\pi_1$ and $P_1$ follow the same set of scheduling rules for the first $k$ classes, this case can only occur when the buffers of streams belonging to the first $k$ classes are either empty or contain non-informative packets for \emph{both} policies. Therefore, (\ref{desired15fork1}) holds naturally. Next, to obtain (\ref{desired15fork2}), we can proceed similarly to Lemma \ref{comparisondelivery} for class $k+1$. The details are therefore omitted for the sake of space.
\end{itemize}
\end{document}